\documentclass[a4paper,11pt]{article}
\usepackage{amsmath, amssymb, amsfonts,indentfirst,graphicx,color,anysize}
\usepackage{lineno}
\marginsize{3cm}{3cm}{2cm}{2cm}
\title{
	\includegraphics[width=0.35\textwidth]{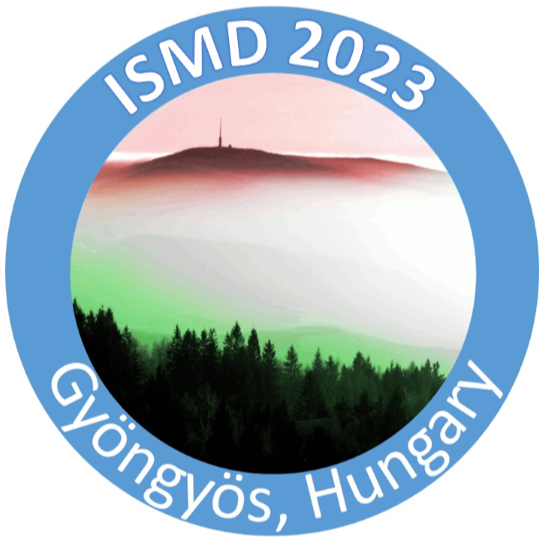}\\[1cm]
	\textbf{Measurements of strange and multi-strange hadrons elliptic flow in isobar collisions at RHIC by STAR}
	}
\author{{V. Bairathi (for the STAR Collaboration)}\\[1ex]
	Instituto de Alta Investigación, Universidad de Tarapacá,\\ Casilla 7D, Arica 1000000, Chile\\
}
\begin{document}

\maketitle

\begin{abstract}
We present measurements of elliptic flow ($v_{2}$) of $K_{s}^{0}$, $\Lambda$, $\bar{\Lambda}$, $\phi$, $\Xi^{-}$, $\overline{\Xi}^{+}$, and $\Omega^{-}$+$\overline{\Omega}^{+}$ at mid-rapidity ($|\eta| <$ 1.0) in isobar collisions ($^{96}_{44}$Ru+$^{96}_{44}$Ru and $^{96}_{40}$Zr+$^{96}_{40}$Zr) at $\sqrt{s_{\mathrm{NN}}}$ = 200 GeV. The centrality and transverse momentum ($p_{\mathrm{T}}$) dependence of elliptic flow is presented. The number of constituent quark (NCQ) scaling of $v_{2}$ in isobar collisions is discussed. $p_{T}$-integrated elliptic flow ($\left\langle v_{2}\right\rangle$) is observed to increase from central to peripheral collisions. The ratio of $\left\langle v_{2}\right\rangle$ between the two isobars shows a deviation from unity for strange hadrons ($K_{s}^{0}$, $\Lambda$ and $\bar{\Lambda}$) indicating a difference in nuclear structure and deformation. A system size dependence of strange hadron $v_{2}$ at high $p_{T}$ is observed among Ru+Ru, Zr+Zr, Cu+Cu, Au+Au, and U+U systems. A multi-phase transport (AMPT) model with string melting (SM) describes the experimental data well in the measured $p_{\mathrm{T}}$ range for isobar collisions at $\sqrt{s_{\mathrm{NN}}}$ = 200 GeV.
\end{abstract}

\section{Introduction}
\label{sec:intro}
Predictions from quantum chromodynamics (QCD) suggest the formation of a de-confined state of quarks and gluons at sufficiently high temperature and energy density called quark-gluon plasma (QGP)~\cite{qgp1,qgp2,qgp3}. Many studies in heavy-ion collisions at the Relativistic Heavy Ion Collider (RHIC)~\cite{rqgp1,rqgp2,rqgp3,rqgp4} and the Large Hadron Collider (LHC)~\cite{aqgp1,aqgp2,aqgp3} have reported the presence of such a medium dominated by the partonic degrees of freedom, which motivates the study of the QGP. The collective flow of produced particles is one of the key observables to probe the QGP medium. It is quantified by the coefficients in the Fourier expansion of azimuthal angle distribution of produced particles with respect to the symmetry planes~\cite{flow1}. This azimuthal anisotropic flow of produced particles indicates hydrodynamic and collective behavior of the strongly interacting matter during the collision~\cite{sqgp1}. It arises due to spatial anisotropy of the initial overlap geometry of the colliding nuclei as a consequence of inhomogeneous initial energy deposition and fluctuations of nucleon positions in heavy-ion collisions. The initial spatial anisotropies are converted into final state momentum anisotropies through multi-particle interactions among partons during the medium evolution. 

The STAR experiment at RHIC collected data in the year 2018 by colliding isobars (Ru+Ru and Zr+Zr) at $\sqrt{s_{\mathrm {NN}}}$ = 200 GeV. It was mainly focused at measuring the charge separation along the magnetic field, driven by a phenomenon called the Chiral Magnetic Effect (CME)~\cite{cme}. The two isobar nuclei have the same atomic mass number but differ in nuclear deformation parameters, and flow measurements are highly sensitive to them. Moreover, the measurement of strange and multi-strange hadrons flow is an excellent probe for understanding the initial state anisotropies due to their small hadronic interaction cross-section compared to light hadrons. Therefore, a systematic study of the anisotropic flow of strange and multi-strange hadrons could be crucial to understanding the effect of initial states in the isobar collisions. 

\section{Dataset and analysis method}
\label{sec:analysis}
In these proceedings, we report $v_{2}$ of $K_{s}^{0}$, $\Lambda$, $\bar{\Lambda}$, $\phi$, $\Xi^{-}$, $\overline{\Xi}^{+}$, and $\Omega^{-}$+$\overline{\Omega}^{+}$ at mid-rapidity ($|\eta| <$ 1.0) in Ru+Ru and Zr+Zr collisions at $\sqrt{s_{\mathrm {NN}}}$ = 200 GeV. A total of $\sim$650 M minimum bias good events for each system out of the total 1.8 B (2 B) events of Ru+Ru (Zr+Zr) collisions are used for this analysis. The above particles are reconstructed using the invariant mass technique through their hadronic decay channels. The combinatorial background is constructed using a rotational method for weakly decaying hadrons, while for $\phi$-mesons event mixing technique is used~\cite{msflow,sflow}. The $\eta$-sub event plane method with a $\eta$-gap of 0.1 between the two sub-events (-1.0 $< \eta <$ -0.05 and 0.05 $< \eta <$ 1.0) is used to calculate $v_{2}$ of these hadrons~\cite{flow1}. The azimuthal dependence of the particle yield can be expanded in terms of a Fourier series with respect to the event plane angle~\cite{flow1}:  
\begin{equation}
E\frac{d^3N}{dp^3} = \frac{d^2N}{2\pi p_{T}dp_{T}dy} \left(1 + 2 \sum_{n=1}^{\infty} v_{n}(p_{T},y) \cos[n(\phi-\Psi_{n})]\right),
\end{equation}
where $p_{T}$ and $y$ are the transverse momentum and rapidity of the particles. The second-order Fourier coefficient $v_{2}$, known as elliptic flow, is particularly sensitive to the initial geometry of the collisions and the properties of the medium in the heavy-ion collisions. $\Psi_{n}$ is the orientation of the n$^{th}$-order event plane. It is reconstructed from the azimuthal distribution of final-state particles as,
\begin{equation}
\psi_{n} = \frac{1}{n} \tan^{-1}\frac{\sum_{i}w_{i}sin(n\phi_{i})}{\sum_{i}w_{i} \cos(n\phi_{i})},
\end{equation}
where, $\phi_{i}$ and $w_{i}$ represent azimuthal angle and weight for the $i^{th}$ particle, respectively. In order to minimize the effects of non-flow correlations, only charged particle tracks with a transverse momentum range of 0.2 $< p_{\mathrm{T}} <$ 2.0 GeV/$c$ are selected to reconstruct the event plane angle. Since the event plane angle is estimated in finite multiplicity, flow coefficients need to be corrected for the event plane resolution. Therefore, the $v_{n}$ measured with respect to the reconstructed event plane is divided by the event plane angle resolution to get the final flow coefficients as,   
\begin{equation}
v_{n} = \frac{v_{n}^{obs}}{\left\langle \cos\left[n(\psi_{n}^{A} - \psi_{n}^{B}\right)] \right\rangle}.
\end{equation}

\begin{figure}[!htbp]
\centering
\includegraphics[width=8cm]{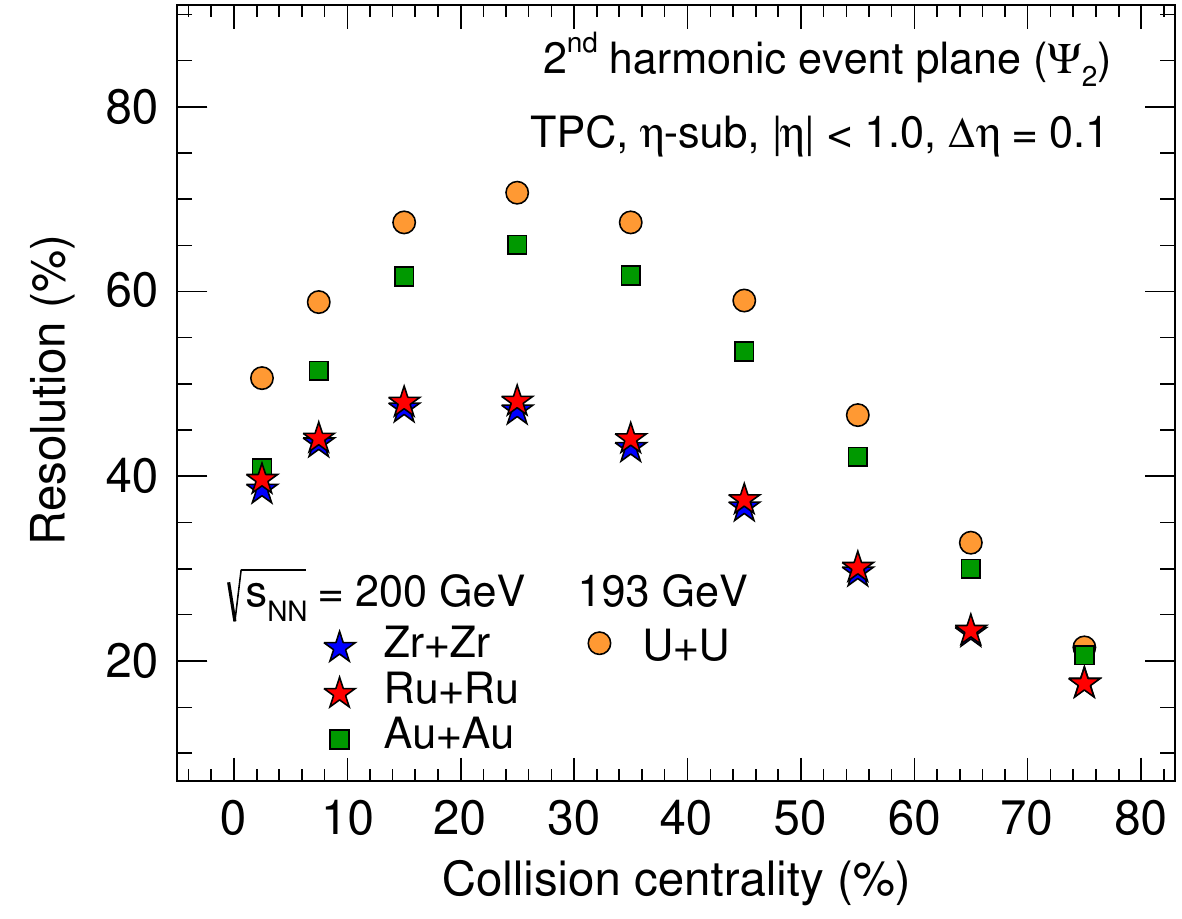}
\caption{2$^{nd}$-order harmonic event plane angle resolution as a function of centrality.}
\label{fig:epres}
\end{figure}

Figure~\ref{fig:epres} shows the event plane resolution as a function of centrality at mid-rapidity ($|\eta| < 1.0$) in Ru+Ru and Zr+Zr collisions at $\sqrt{s_{\mathrm {NN}}}$ = 200 GeV. For comparison, event plane resolution from the published results in Au+Au collisions at $\sqrt{s_{\mathrm {NN}}}$ = 200 GeV and U+U collisions at $\sqrt{s_{\mathrm {NN}}}$ = 193 GeV are also shown. The event plane resolution is better for systems with more multiplicity and the number of participants. It shows similar centrality dependence in all the systems.

\section{Results}
\label{sec:results}
\subsection{$p_{T}$ dependence of $v_{2}$}
\label{ssec:v2pt}
\begin{figure}[!htbp]
\centering
\includegraphics[width=6.5cm]{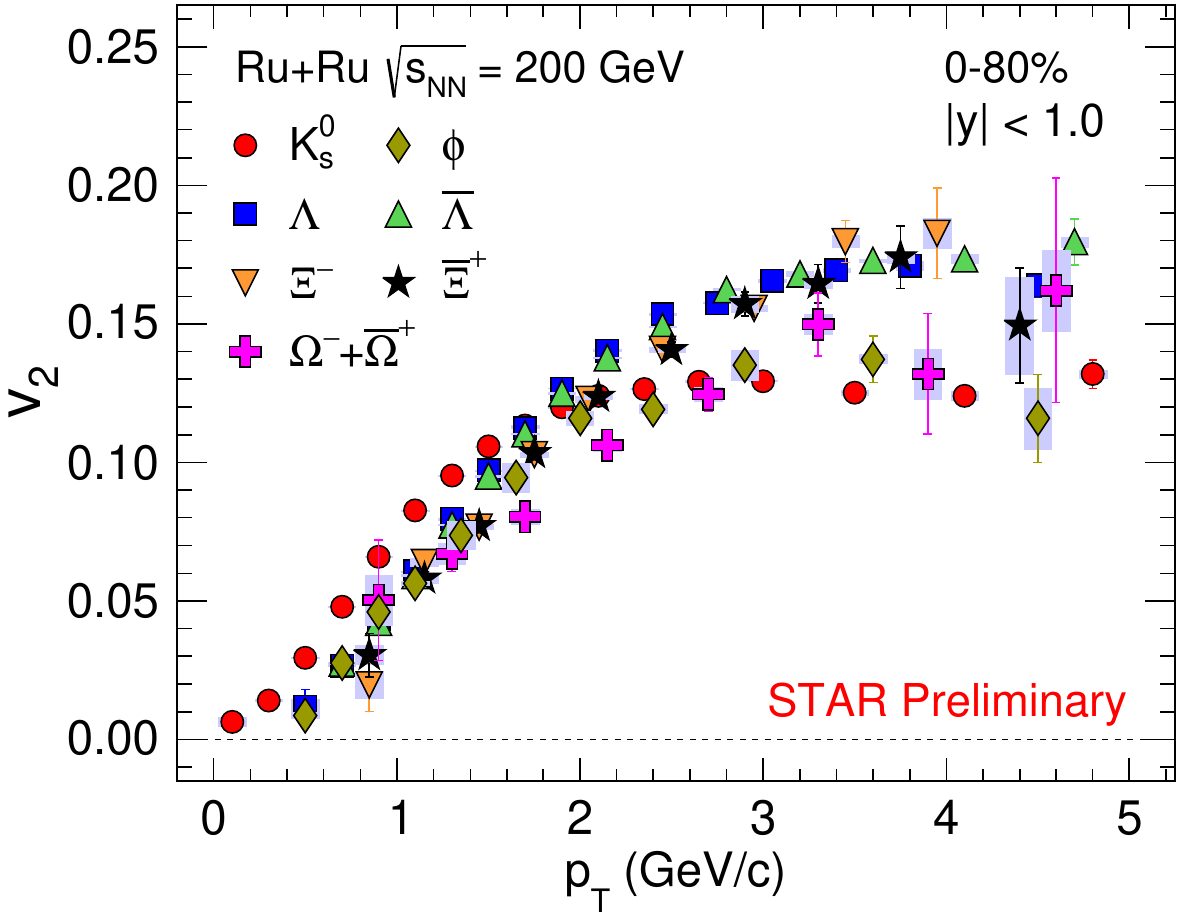}
\includegraphics[width=6.5cm]{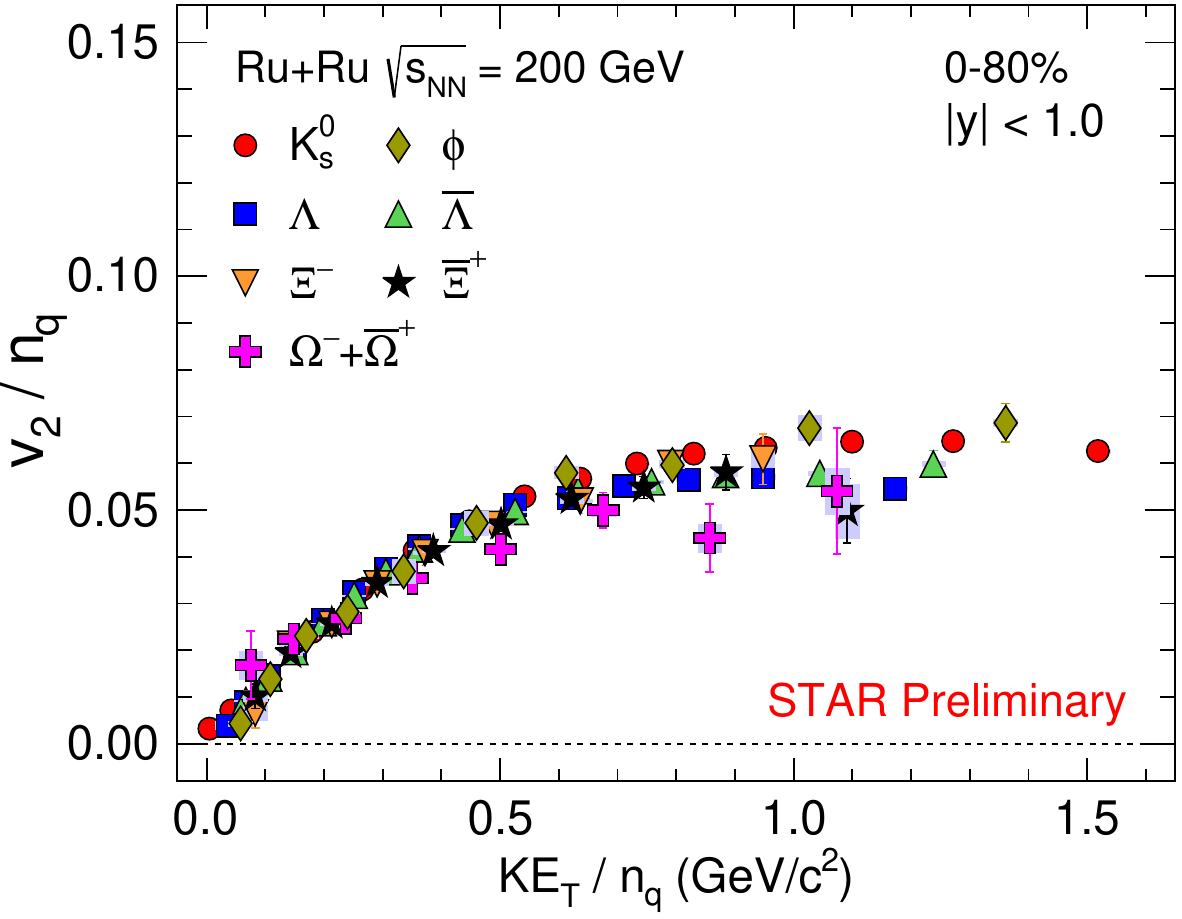}
\includegraphics[width=6.5cm]{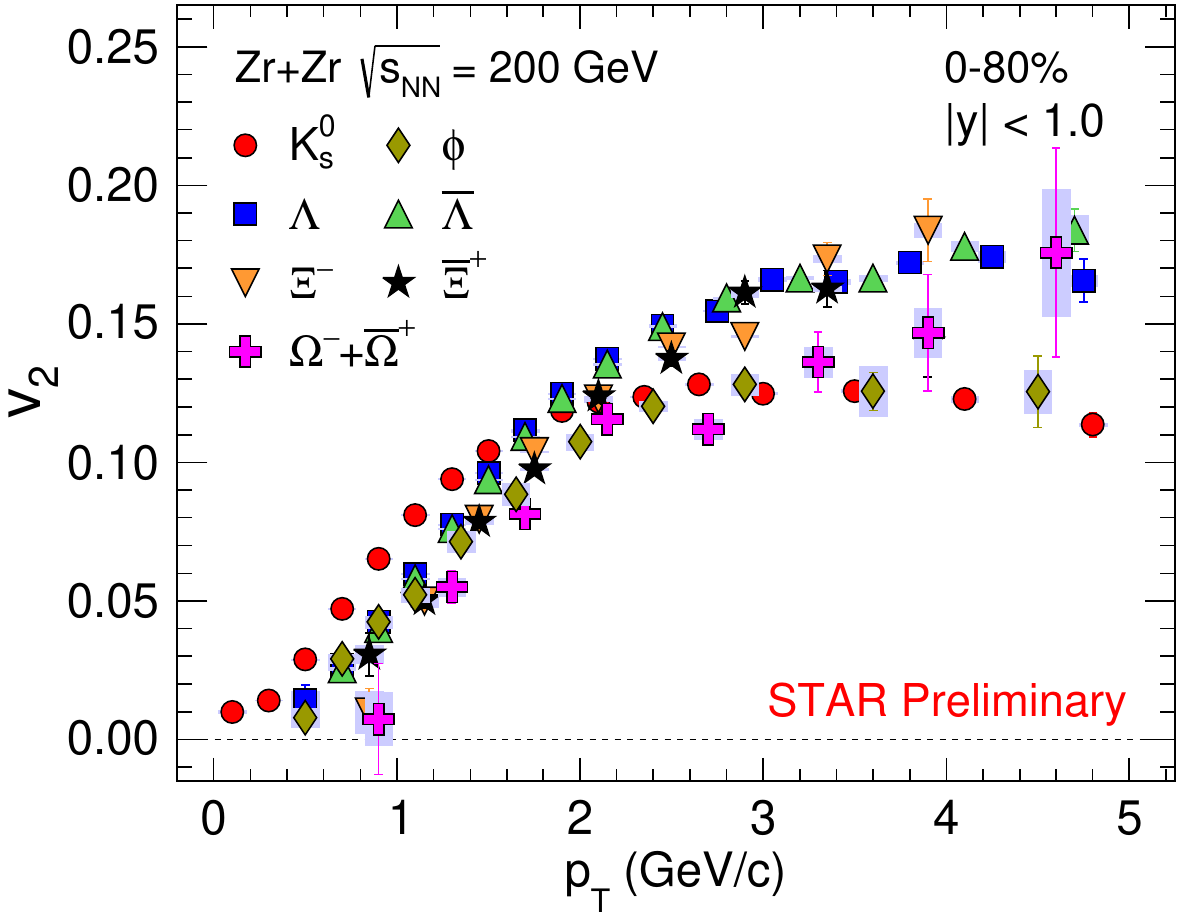}
\includegraphics[width=6.5cm]{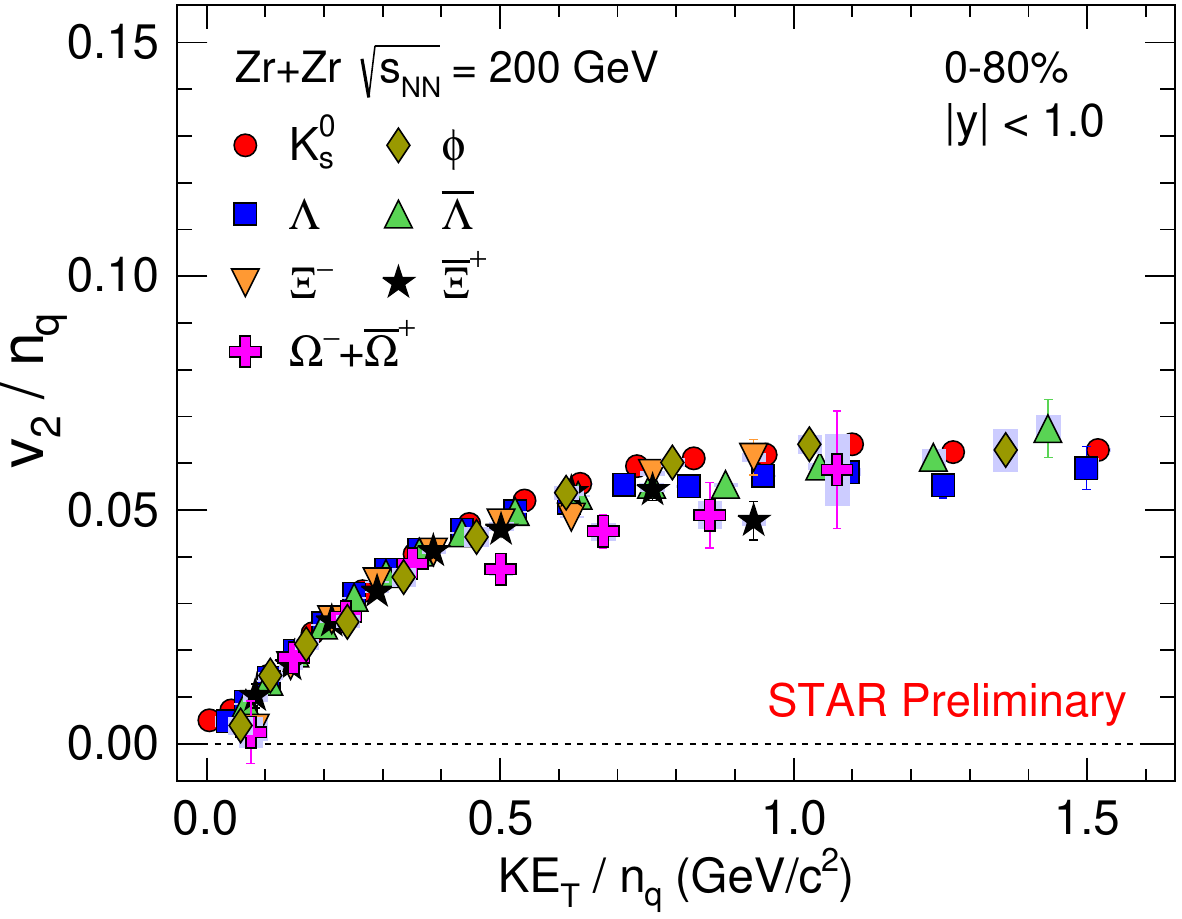}
\caption{$v_{2}$ as a function of $p_{T}$ for $K_{s}^{0}$, $\Lambda$, $\bar{\Lambda}$, $\phi$, $\Xi^{-}$, $\overline{\Xi}^{+}$, and $\Omega^{-}$+$\overline{\Omega}^{+}$ at mid-rapidity in minimum bias Ru+Ru collisions (top left panel) and Zr+Zr collisions (bottom left panel) at $\sqrt{s_{\mathrm {NN}}}$ = 200 GeV. NCQ scaled $v_{2}$ vs transverse kinetic energy ($KE_{T}$/$n_{q}$) is also shown for Ru+Ru collisions (top right panel) and Zr+Zr collisions (bottom right panel) at $\sqrt{s_{\mathrm {NN}}}$ = 200 GeV. The bands represent systematic uncertainties.}
\label{fig:v2mb}
\end{figure}

Figure~\ref{fig:v2mb} shows strange and multi-strange hadrons $v_{2}$ as function of $p_{T}$ for minimum bias (0-80\%) Ru+Ru and Zr+Zr collisions at $\sqrt{s_{\mathrm {NN}}}$ = 200 GeV. $v_{2}$ follows a particle mass ordering indicating hydrodynamic behavior of the medium at low $p_{T}$. Whereas, at intermediate $p_{T}$, it shows a splitting between baryons and mesons, which suggests the formation of QGP medium in isobar collisions at $\sqrt{s_{\mathrm {NN}}}$ = 200 GeV. A similar transverse momentum dependence of $v_{2}$ is observed for both Ru+Ru and Zr+Zr collisions.

Figure~\ref{fig:v2mb} also shows $v_{2}$ of strange and multi-strange hadrons scaled by the number of constituent quarks $n_{q}$ in minimum bias Ru+Ru and Zr+Zr collisions at $\sqrt{s_{\mathrm {NN}}}$ = 200 GeV. The results are presented as a function of transverse kinetic energy to remove the effect of particle mass at low $p_{T}$. It is defined as $KE_{T} = m_{T} - m_{0}$, where $m_{T}$ is the transverse mass ($\sqrt{p_{T}^{2}+m_{0}^{2}}$) and $m_{0}$ is rest mass of the particle. The $v_{2}$ of strange and multi-strange hadrons follows the number of constituent quarks (NCQ) scaling with $\pm$10\% uncertainty in both collision systems. The NCQ scaling of $v_{2}$ suggests the formation and collective behavior of the QGP medium. It also indicates that quark coalescence is the dominant mechanism of particle production. 

\subsection{Centrality dependence of $v_{2}$}
\label{ssec:v2cent}
\begin{figure}[!htbp]
\centering
\includegraphics[width=4.9cm]{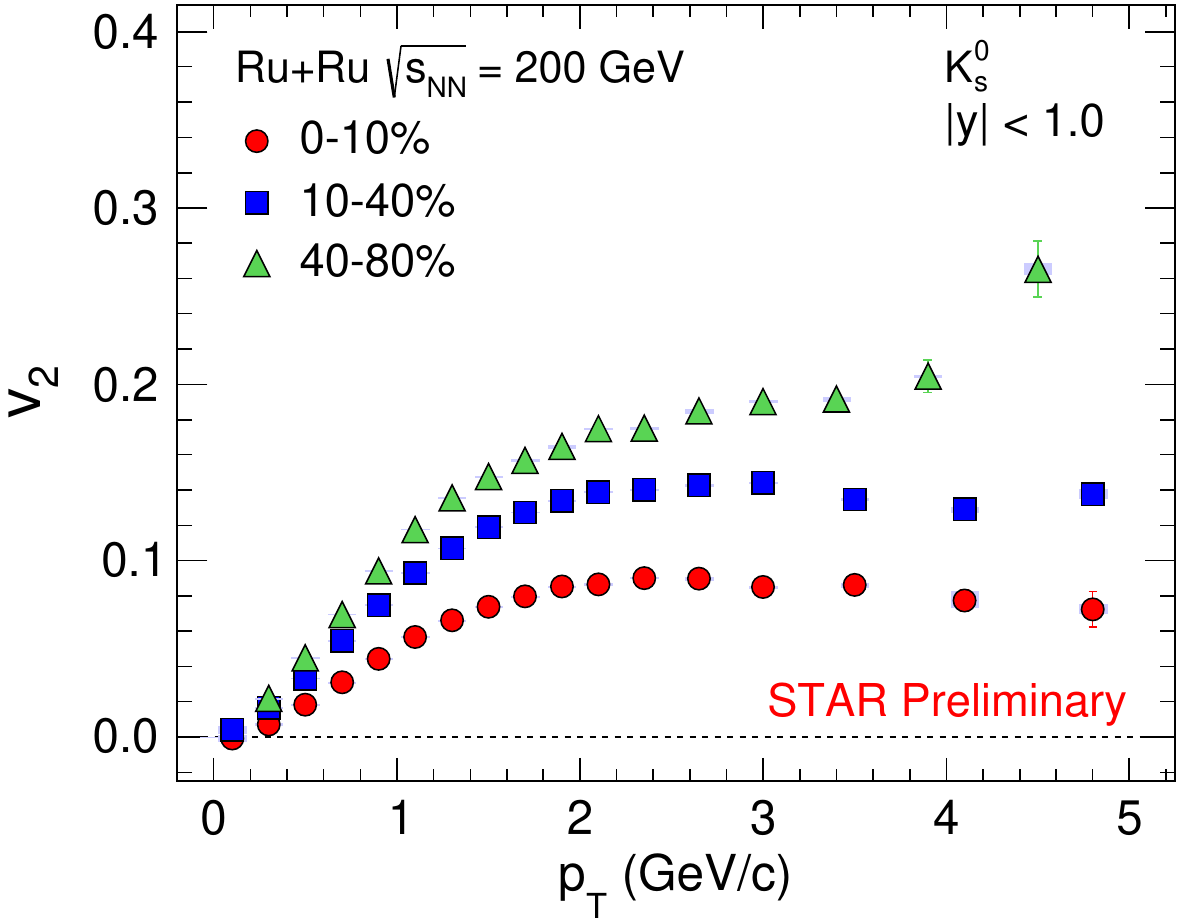}
\includegraphics[width=4.9cm]{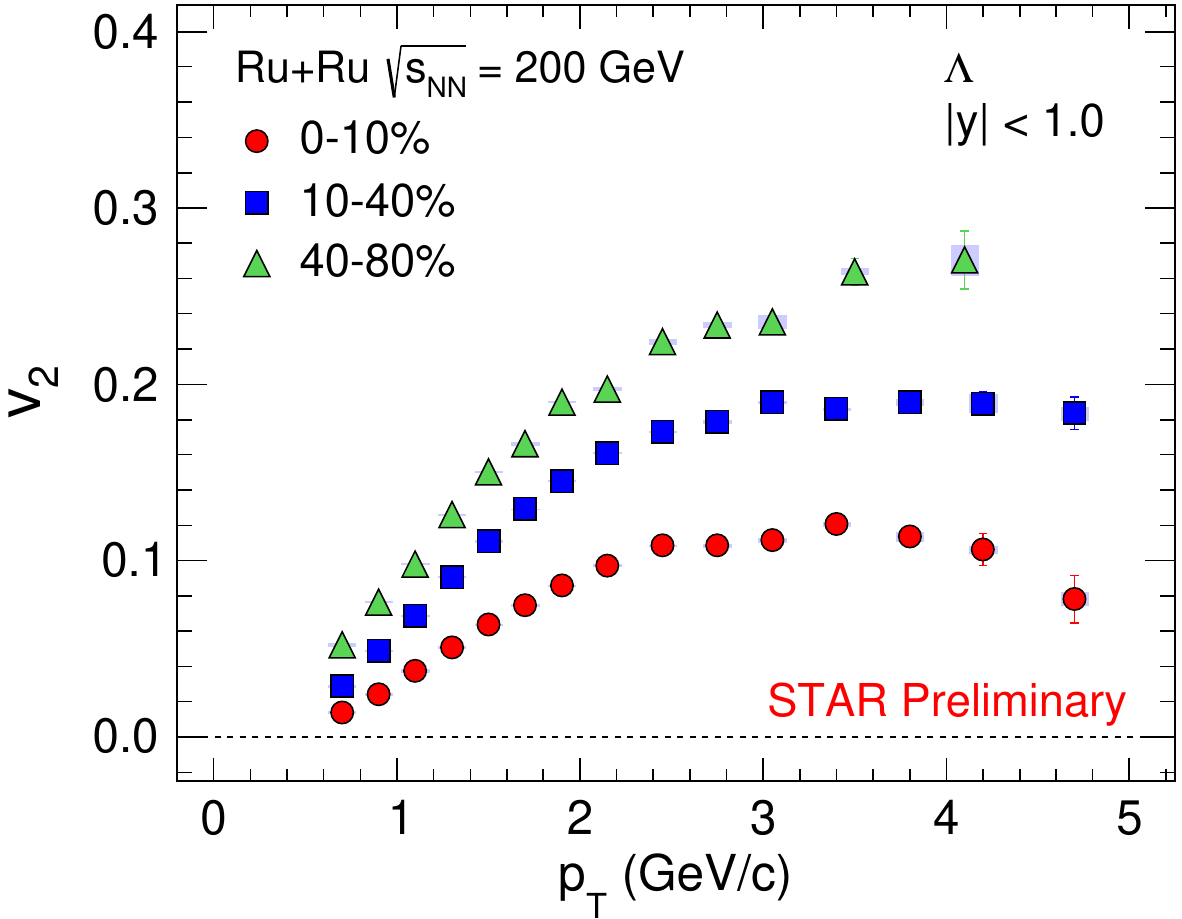}
\includegraphics[width=4.9cm]{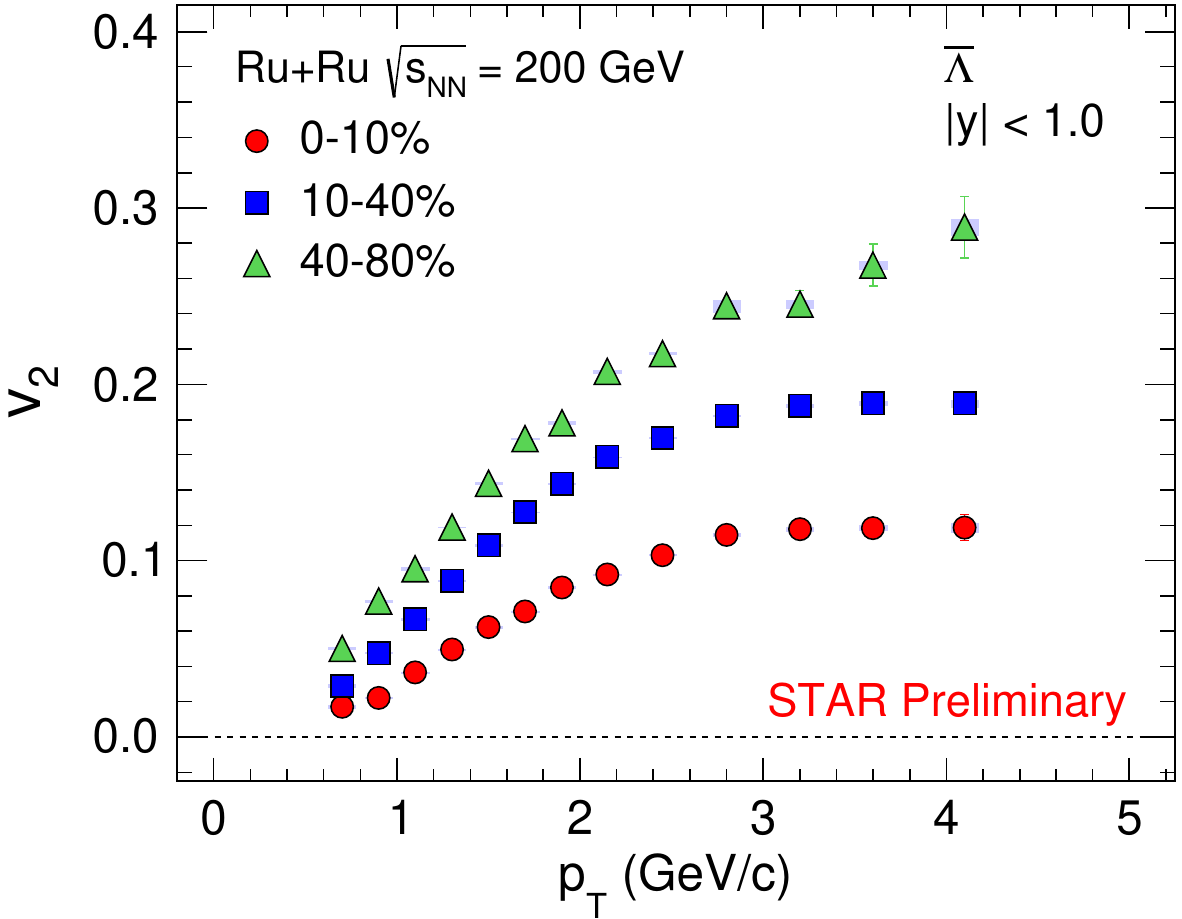}
\includegraphics[width=4.9cm]{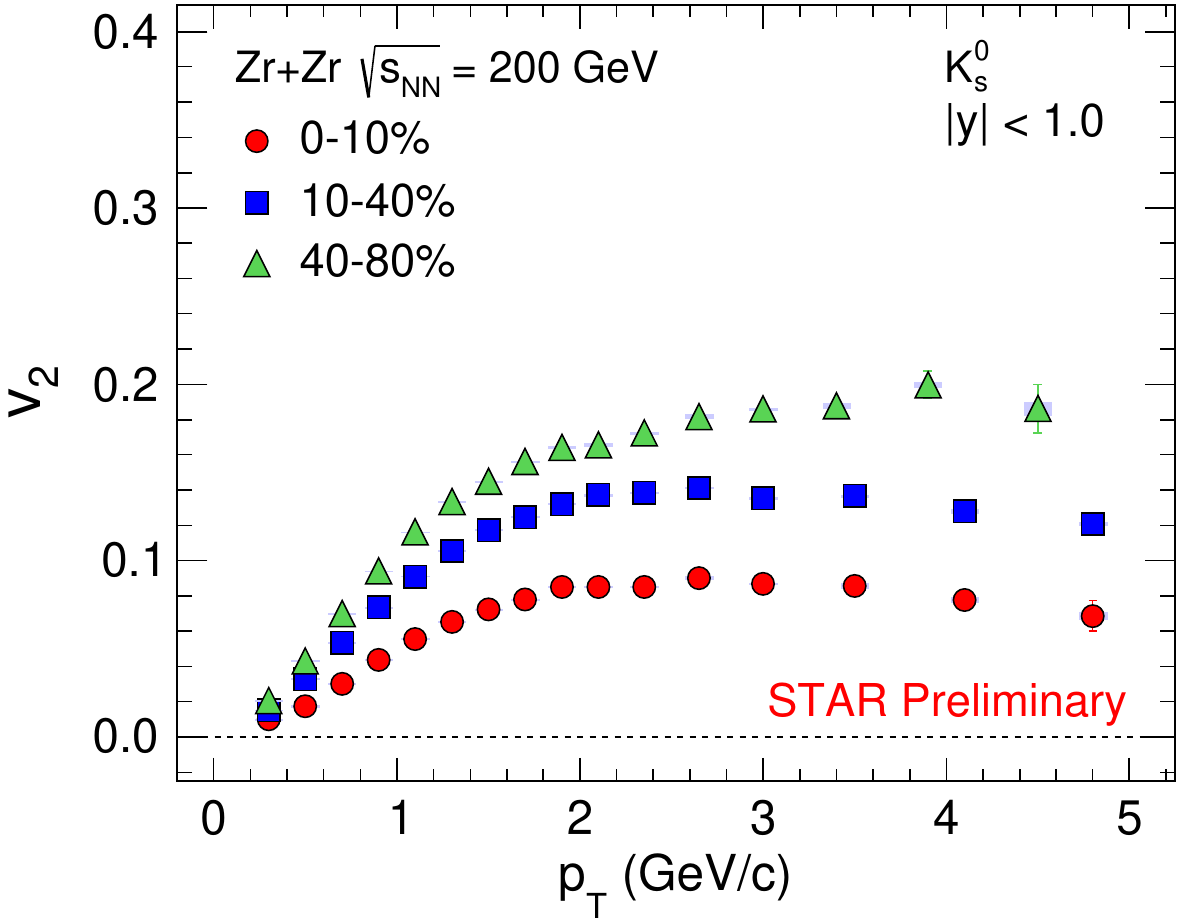}
\includegraphics[width=4.9cm]{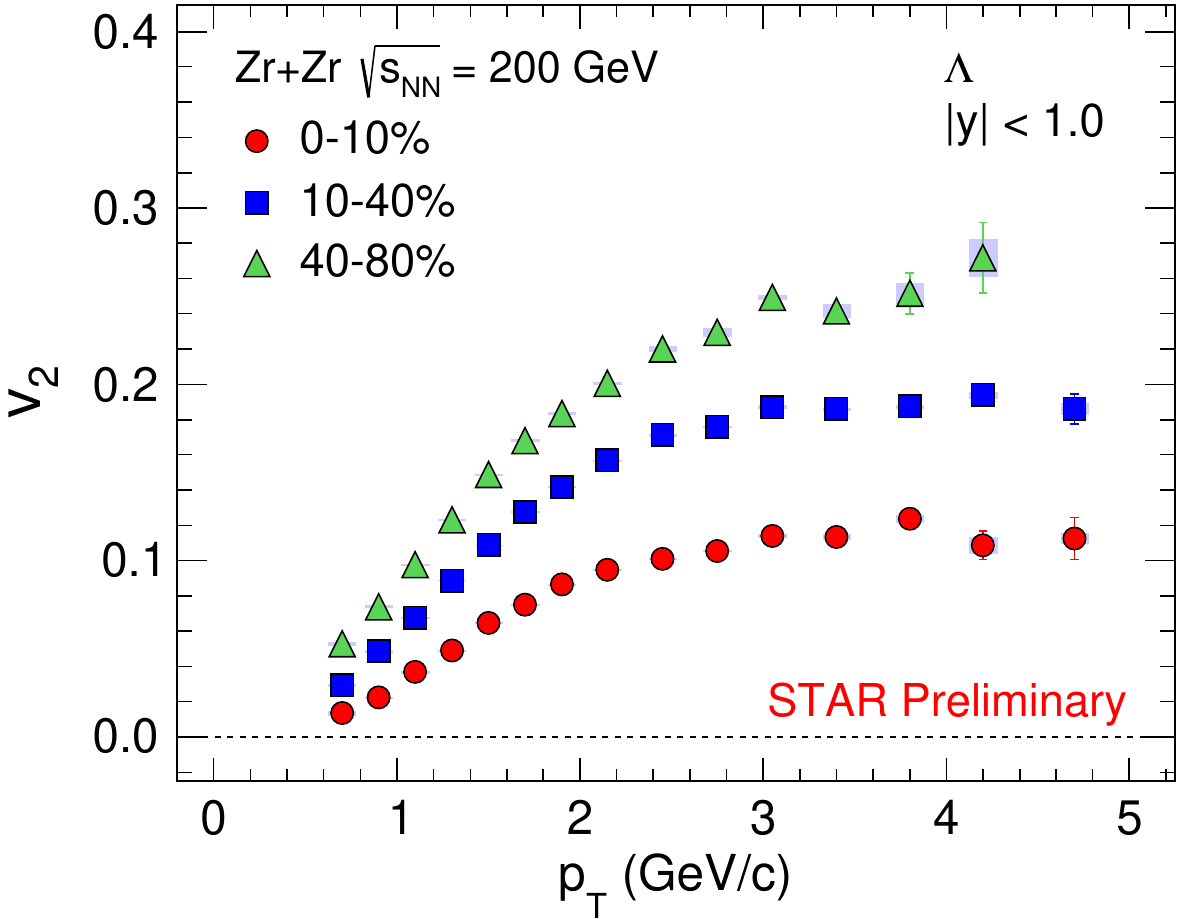}
\includegraphics[width=4.9cm]{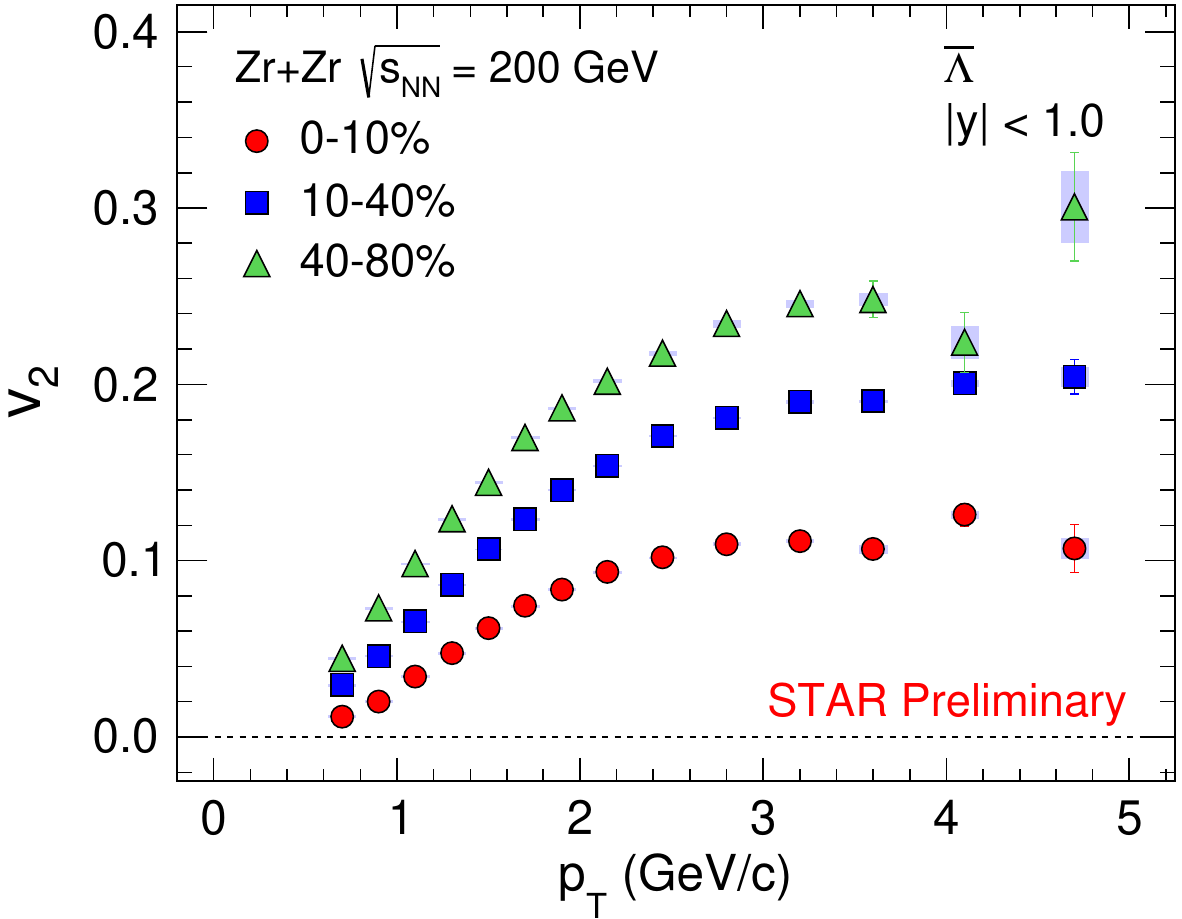}
\caption{$v_{2}$($p_{T}$) of strange hadrons at mid-rapidity in Ru+Ru (top panels) and Zr+Zr (bottom panels) collisions at $\sqrt{s_{\mathrm {NN}}}$ = 200 GeV for centrality 0-10\%, 10-40\%, and 40-80\%. The bands represent systematic uncertainties.}
\label{fig:v2cent_str}
\end{figure}
\begin{figure}[!htbp]
\centering
\includegraphics[width=4.9cm]{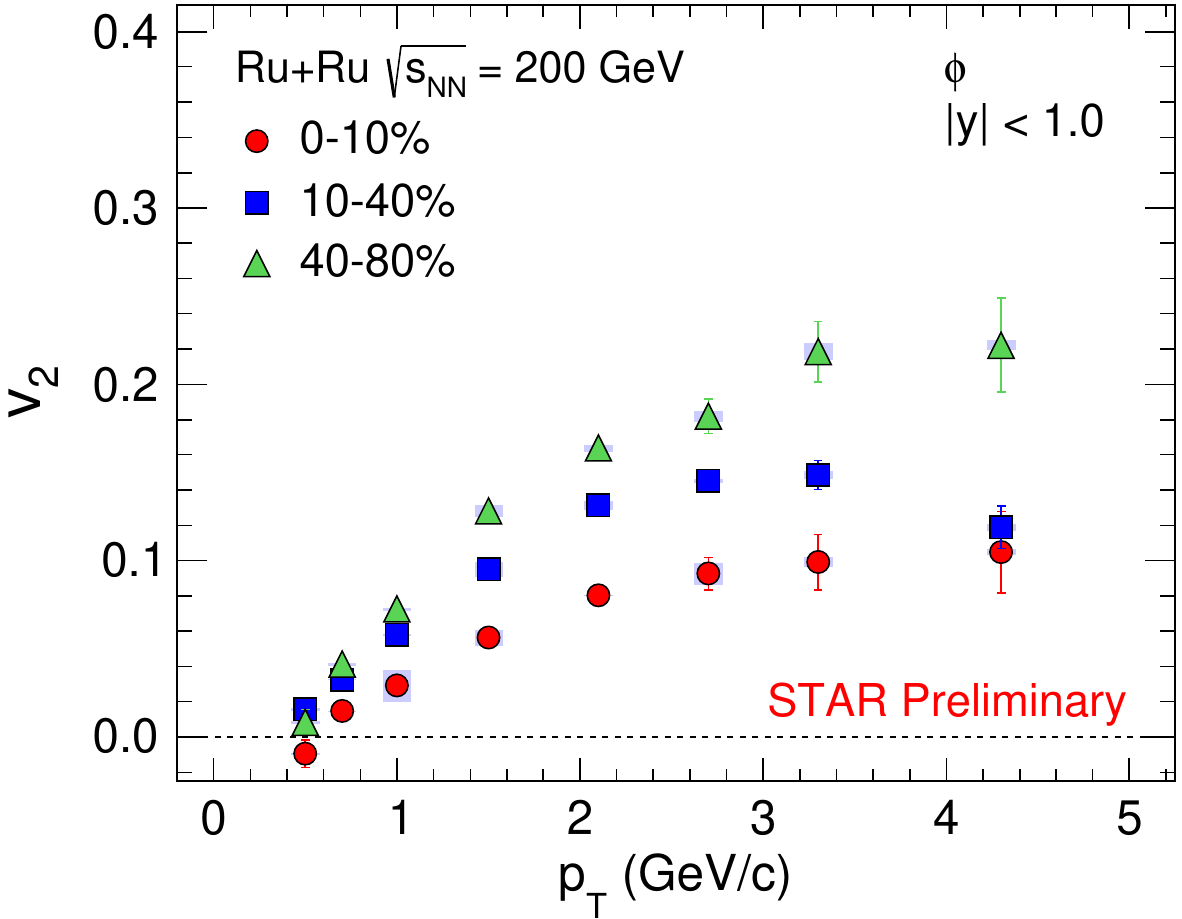}
\includegraphics[width=4.9cm]{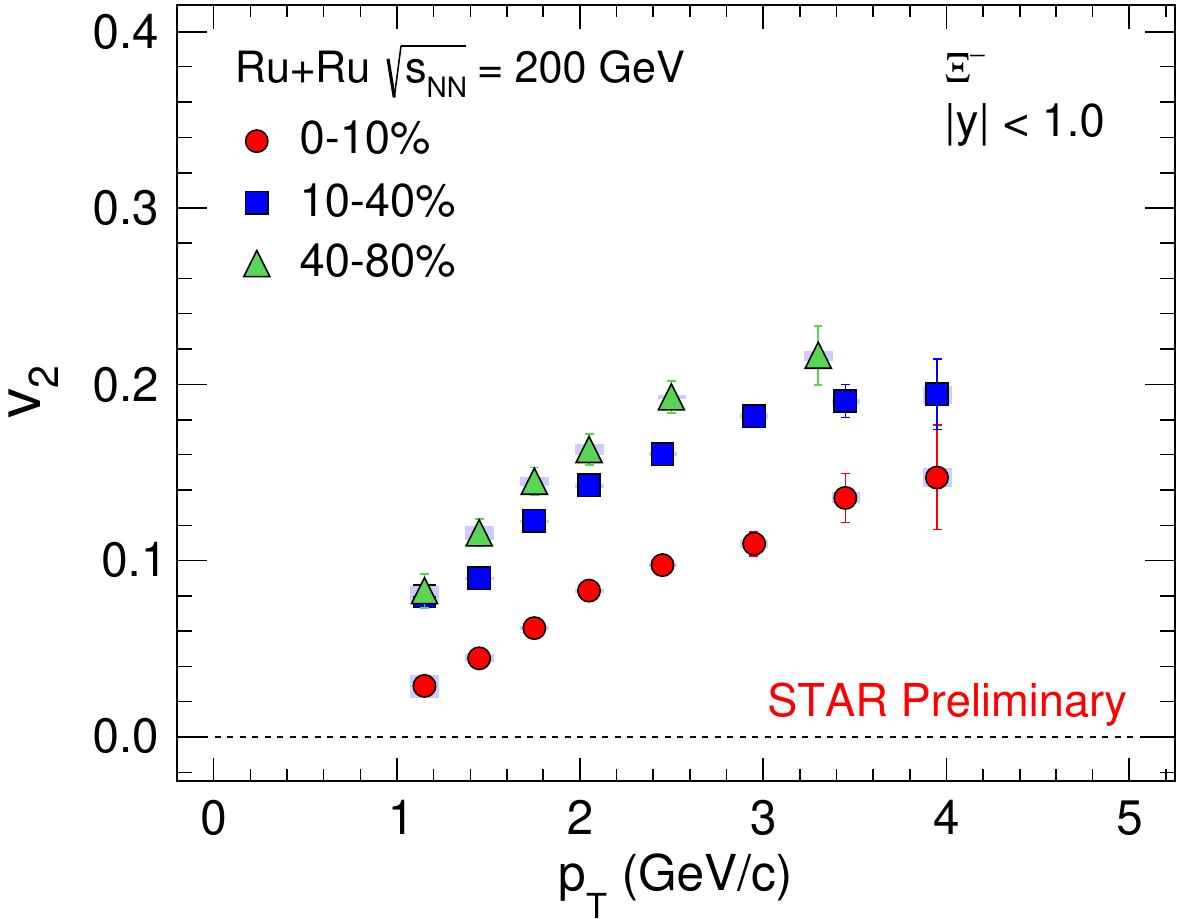}
\includegraphics[width=4.9cm]{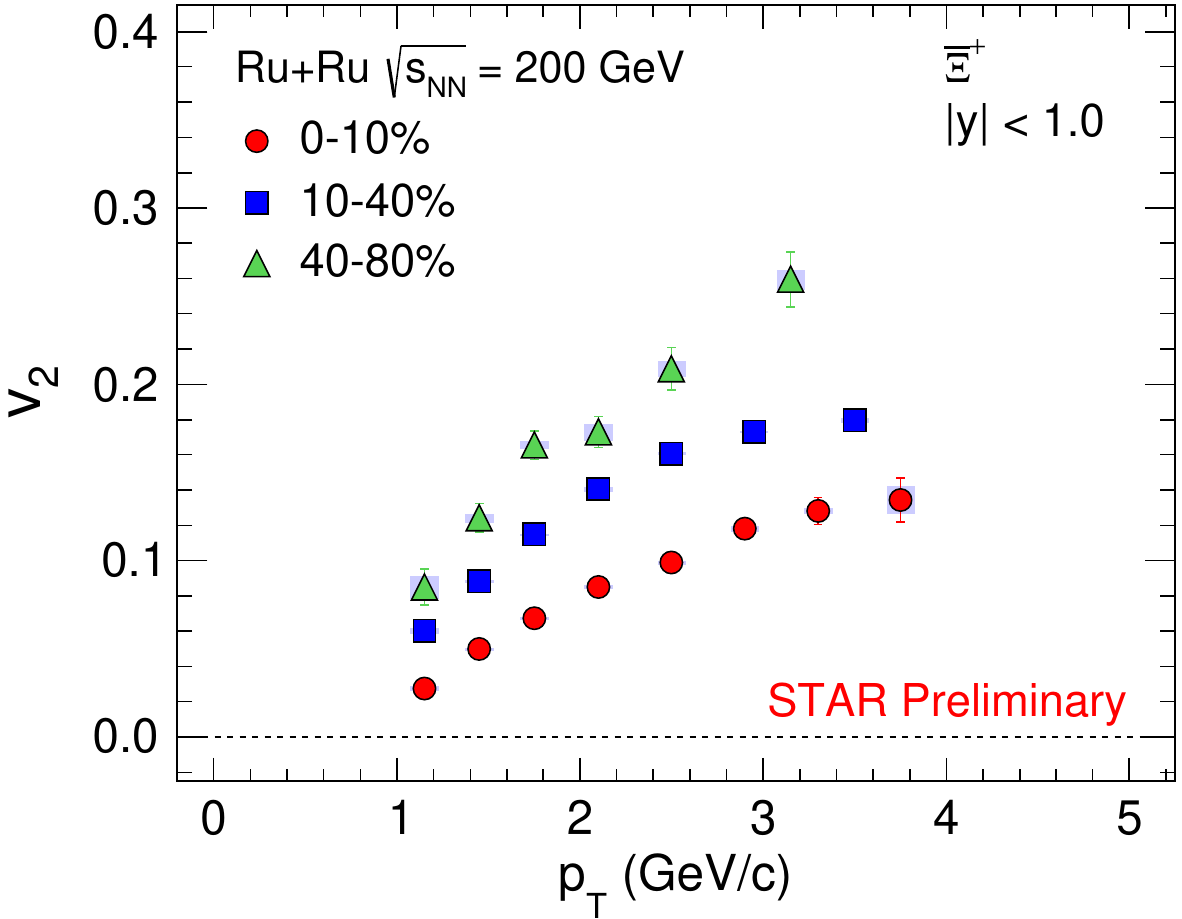}
\includegraphics[width=4.9cm]{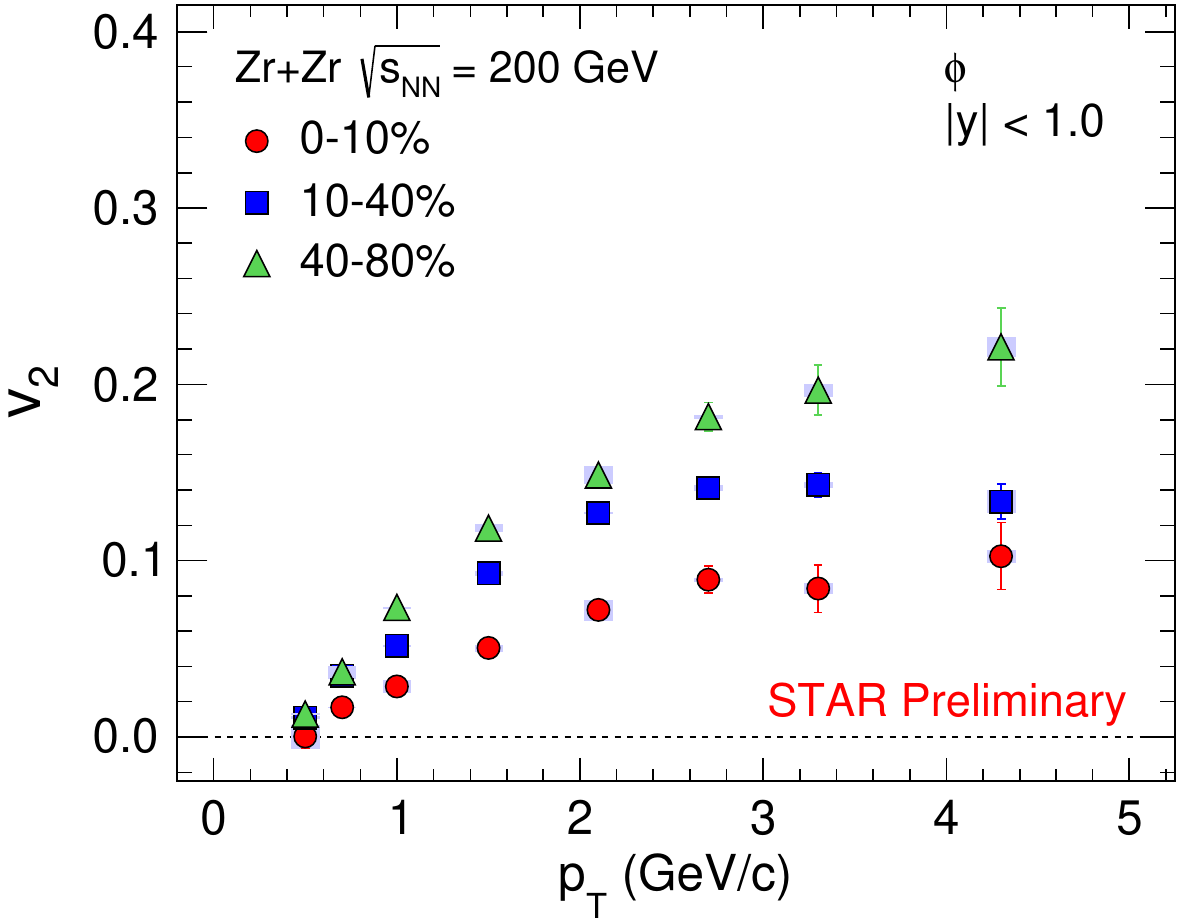}
\includegraphics[width=4.9cm]{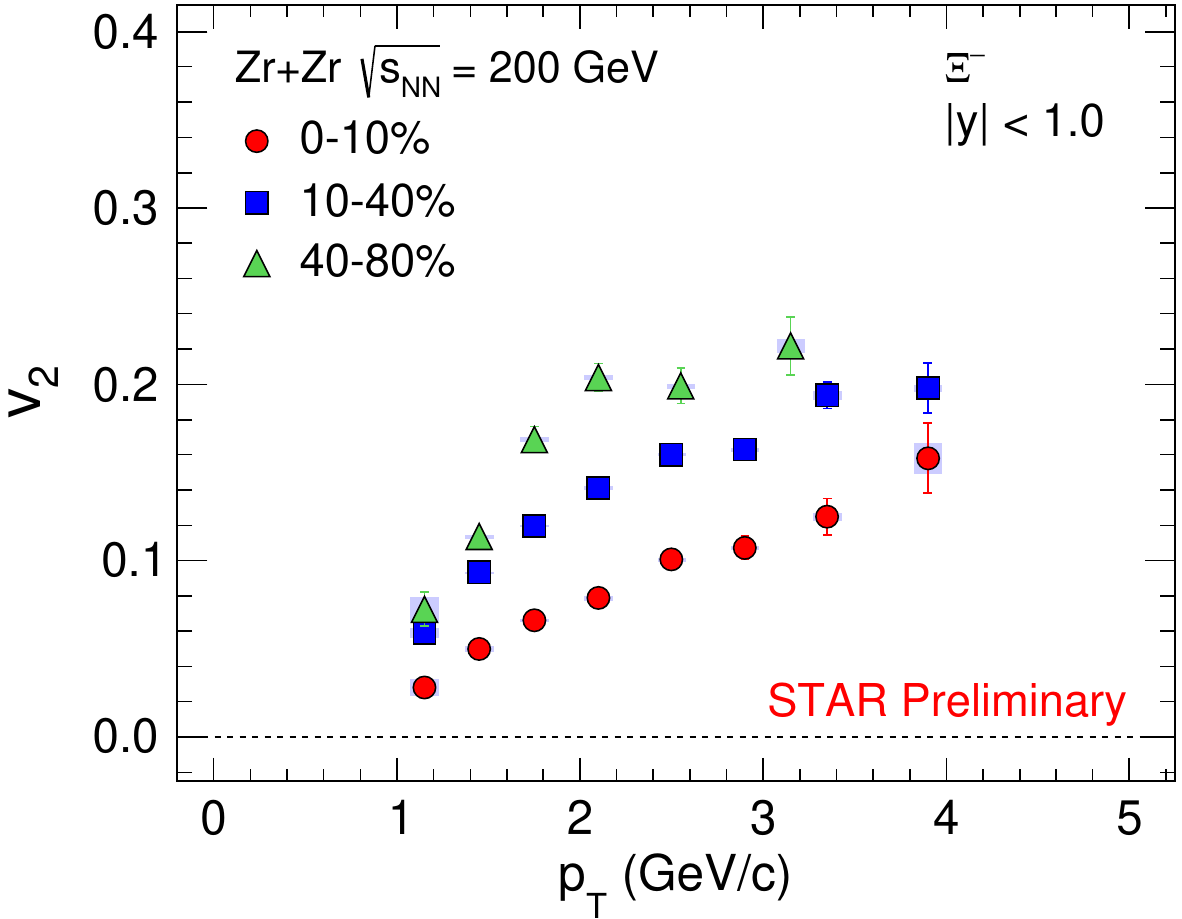}
\includegraphics[width=4.9cm]{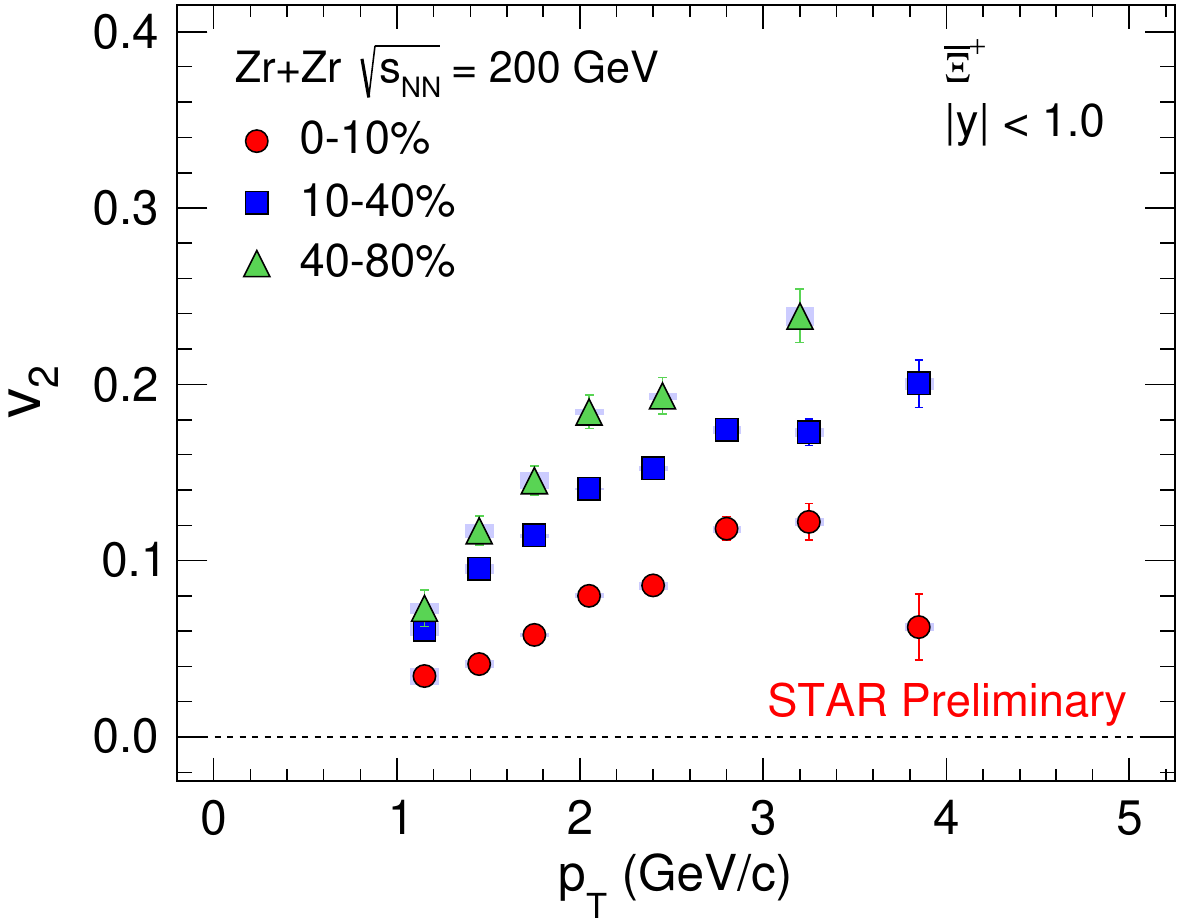}
\caption{$v_{2}$($p_{T}$) of multi-strange hadrons at mid-rapidity in Ru+Ru (top panels) and Zr+Zr (bottom panels) at $\sqrt{s_{\mathrm {NN}}}$ = 200 GeV for centrality 0-10\%, 10-40\%, and 40-80\%. The bands represent systematic uncertainties.}
\label{fig:v2cent_ms}
\end{figure}
Figures~\ref{fig:v2cent_str} and~\ref{fig:v2cent_ms} show $v_{2}(p_{T})$ of strange and multi-strange hadrons for various centrality intervals in Ru+Ru and Zr+Zr collisions at $\sqrt{s_{\mathrm {NN}}}$ = 200 GeV. A strong centrality dependence is observed for all the particles studied in both the isobar systems. The magnitude of $v_{2}$ increases from central (0-10\%) to peripheral (40-80\%) collisions, which indicate the effect of initial eccentricity in isobar collisions at $\sqrt{s_{\mathrm {NN}}}$ = 200 GeV.
\begin{figure}[!htbp]
\centering
\includegraphics[width=4.9cm]{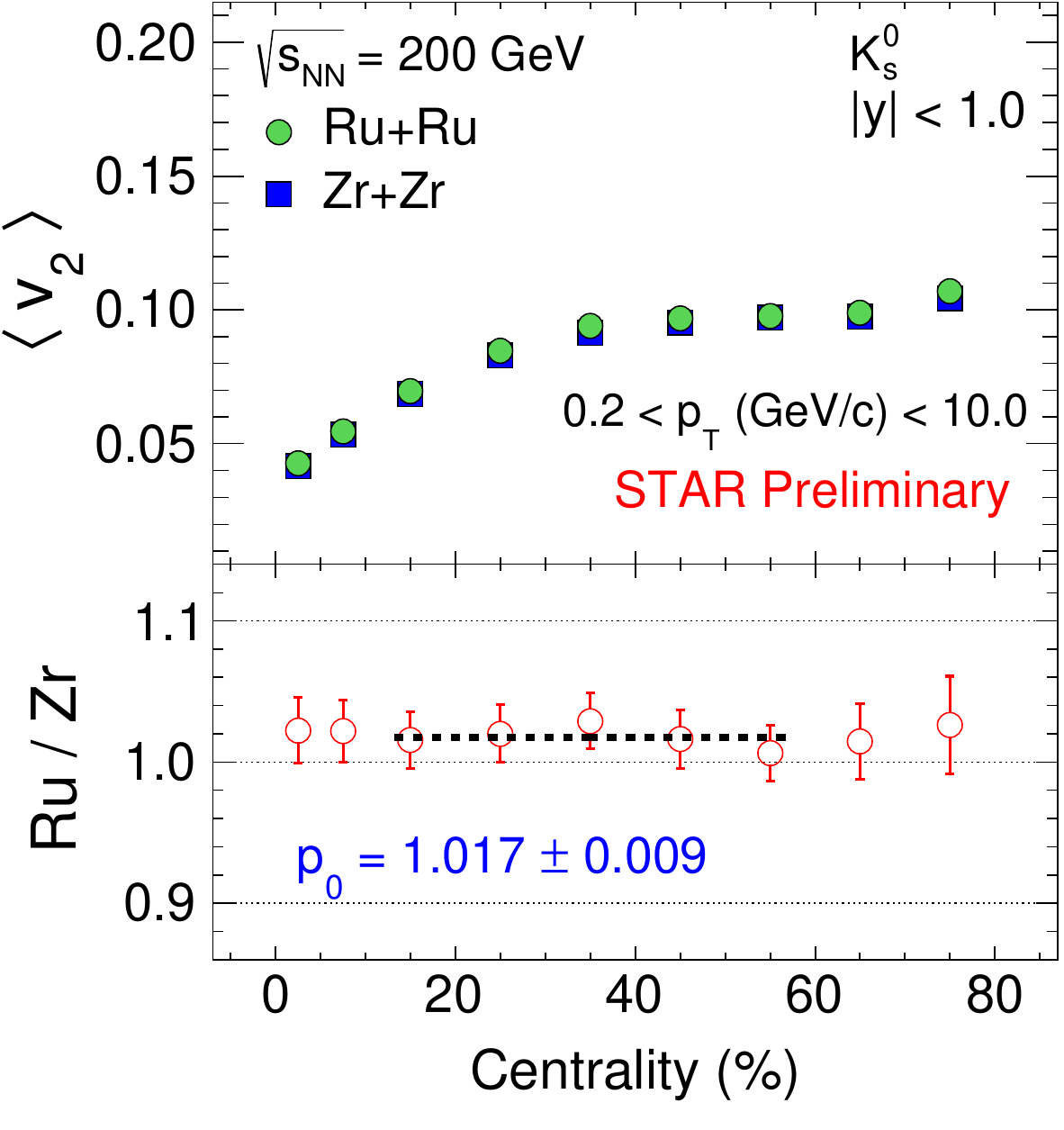}
\includegraphics[width=4.9cm]{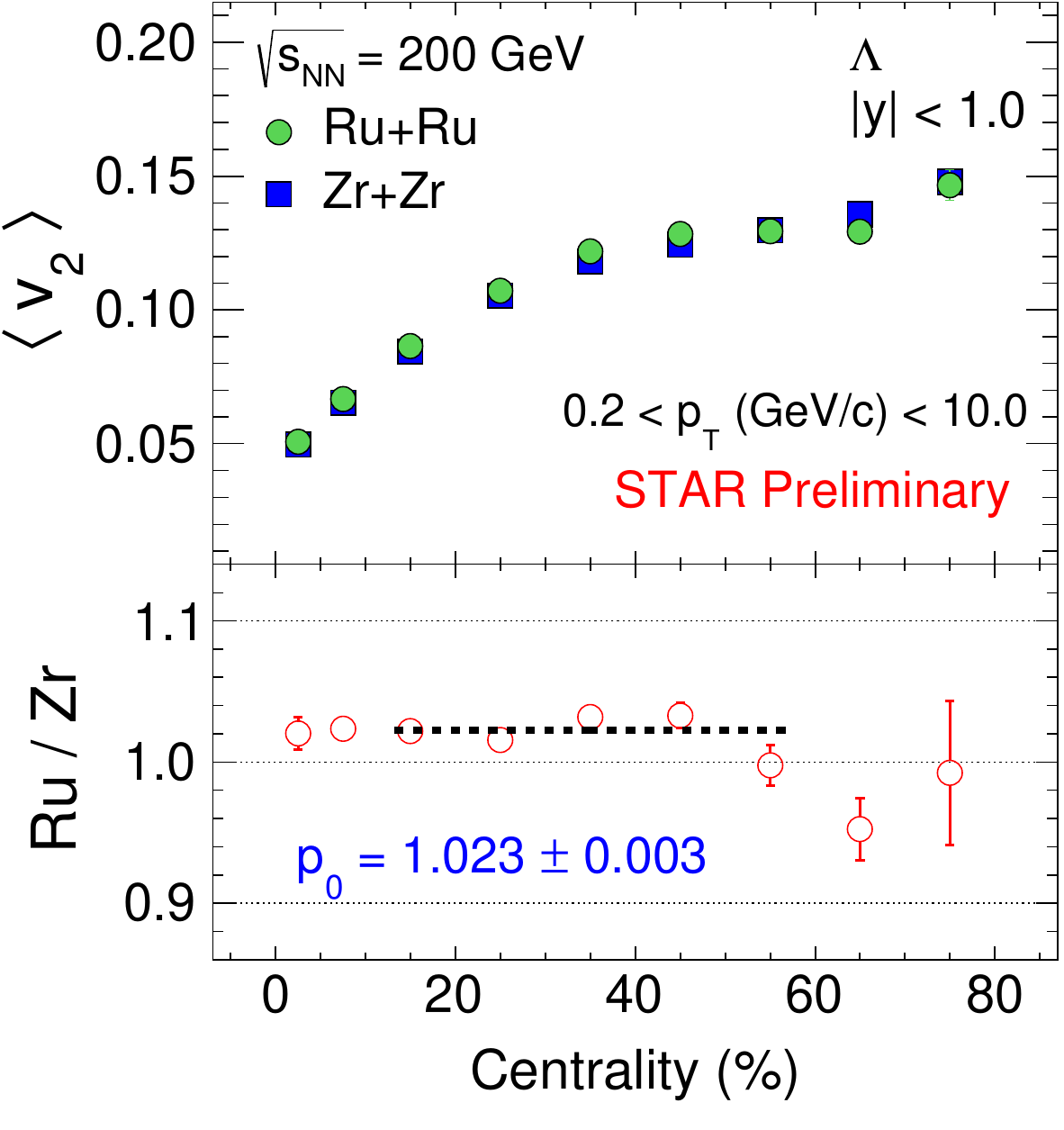}
\includegraphics[width=4.9cm]{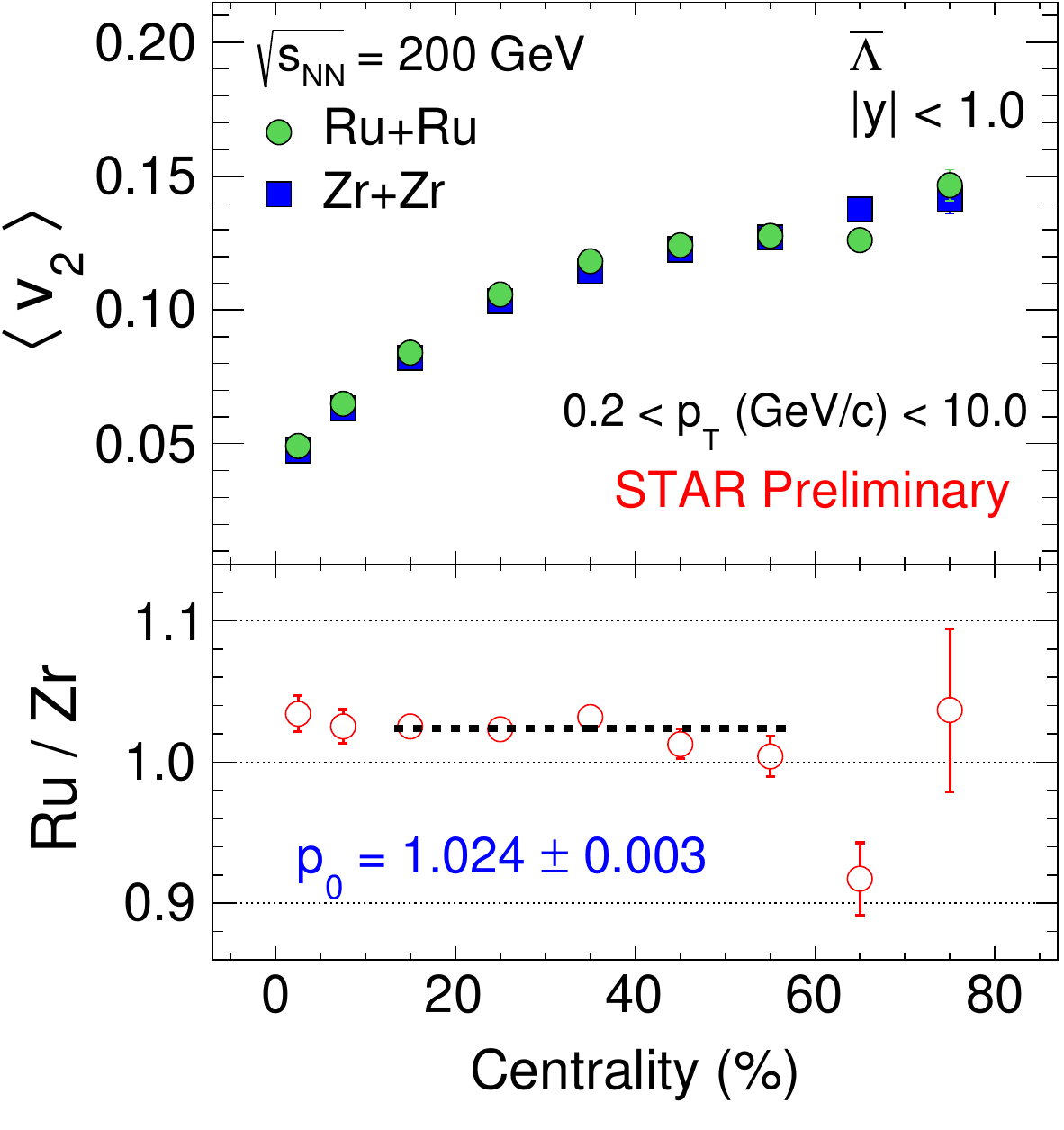}
\caption{$p_{T}$-integrated $v_{2}$ vs centrality for strange hadrons at mid-rapidity in Ru+Ru and Zr+Zr collisions at $\sqrt{s_{\mathrm {NN}}}$ = 200 GeV. The bottom panels also show the ratio of $v_{2}$ between Ru and Zr. The error bars represent statistical and systematic uncertainties added in quadrature.}
\label{fig:v2cent_ratio}
\end{figure}

Figure~\ref{fig:v2cent_ratio} shows $p_{T}$-integrated $v_{2}$ of strange hadrons as a function of centrality in Ru+Ru and Zr+Zr collisions at $\sqrt{s_{\mathrm {NN}}}$ = 200 GeV. The ratio of $v_{2}$ in Ru+Ru to Zr+Zr collisions is also shown in the bottom panels of Fig.~\ref{fig:v2cent_ratio} and fitted with a constant polynomial function for mid-central collisions (20-50\%). About $\sim$2\% deviation from unity with a significance of 6.25$\sigma$ for $\Lambda(\bar{\Lambda})$ and 1.83$\sigma$ for $K_{s}^{0}$ is observed, which is consistent with the expectation from the difference between the nuclear structures of the two isobar nuclei~\cite{isobarData}.

\subsection{System size dependence}
\label{ssec:v2sys}
\begin{figure}[!htbp]
\centering
\includegraphics[width=4.9cm]{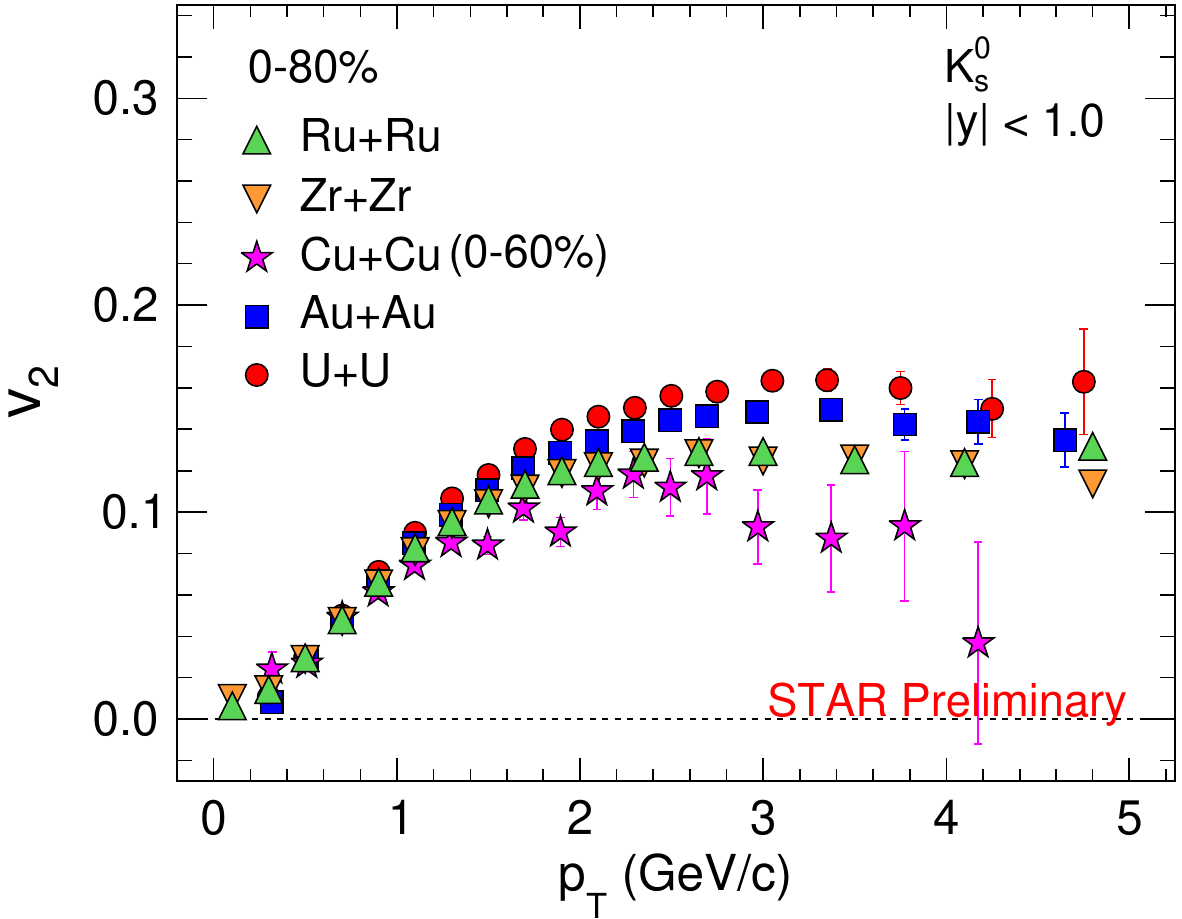}
\includegraphics[width=4.9cm]{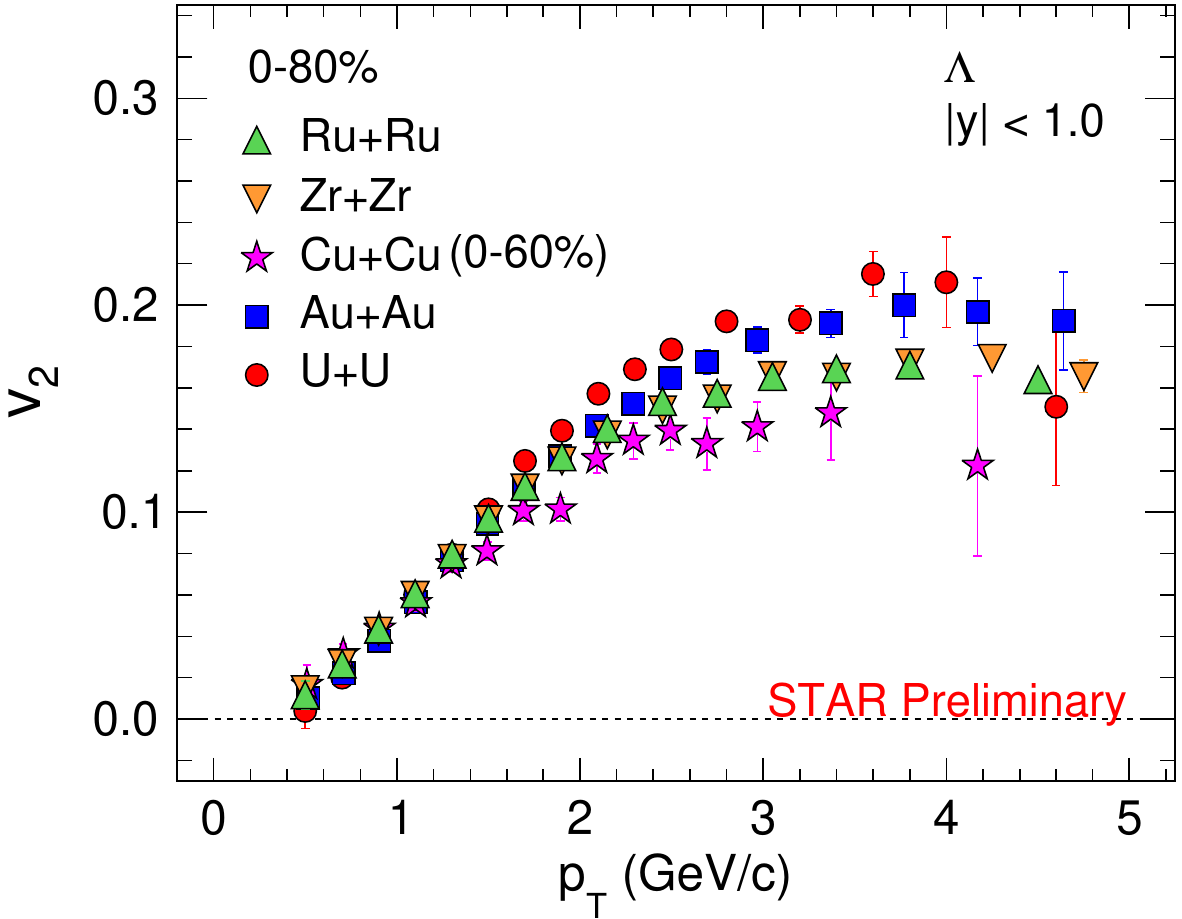}
\includegraphics[width=4.9cm]{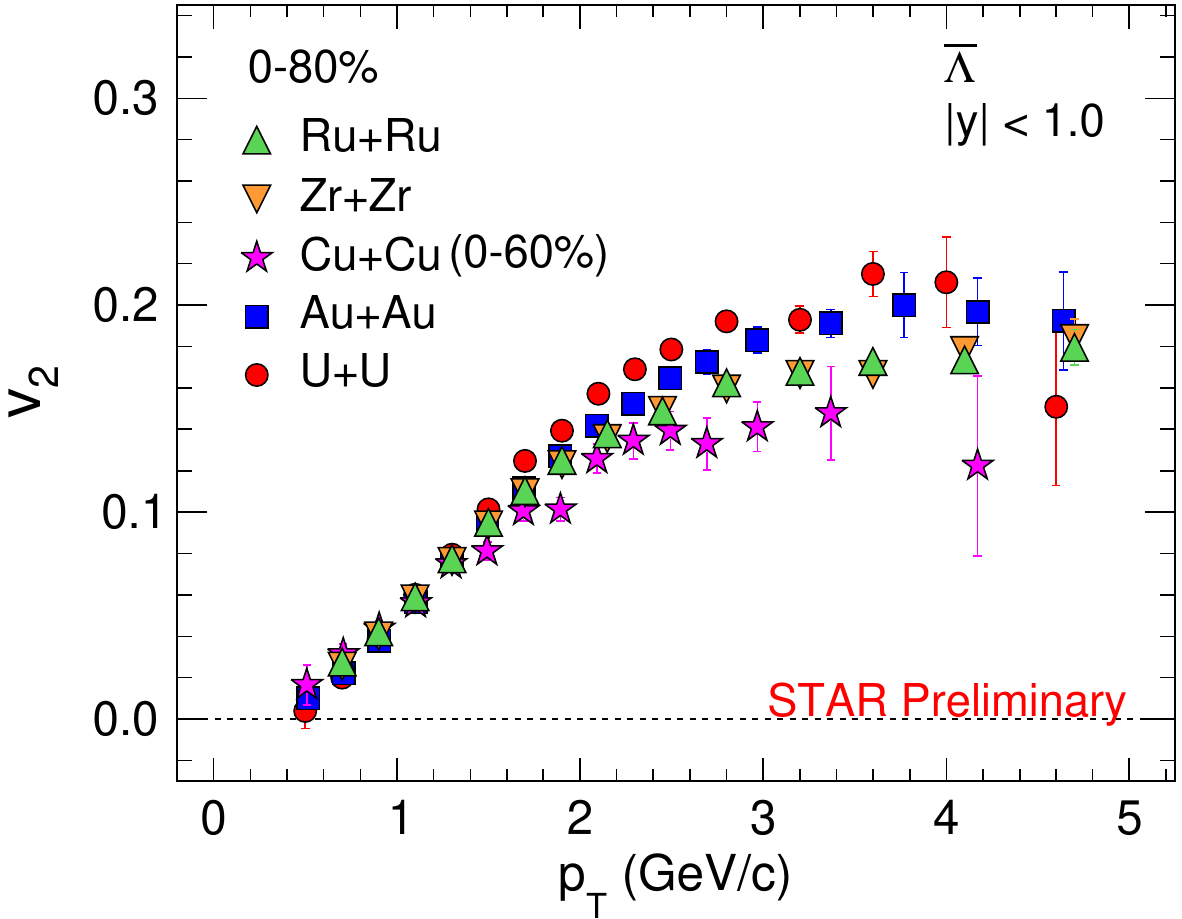}
\caption{Strange hadron $v_{2}$ as a function of $p_{T}$ at mid-rapidity in minimum bias Ru+Ru and Zr+Zr collisions at $\sqrt{s_{\mathrm {NN}}}$ = 200 GeV compared to Cu+Cu, Au+Au, and U+U collisions~\cite{auau,cucu,uu}. The error bars represent statistical and systematic uncertainties added in quadrature.}
\label{fig:v2sys}
\end{figure}

Figure~\ref{fig:v2sys} shows $v_{2}$ of strange hadrons in Ru+Ru and Zr+Zr collisions at $\sqrt{s_{\mathrm {NN}}}$ = 200 GeV compared to the published results from the STAR experiment at RHIC in Cu+Cu, Au+Au, and U+U collisions~\cite{auau,cucu,uu}. A system size dependence of $v_{2}$ is observed for $p_{T}$ above $\sim$1.5 GeV/$c$. The $v_{2}$ follow the hierarchy $v_{2}^{\textrm{Cu}} < v_{2}^{\textrm{Ru/Zr}} < v_{2}^{\textrm{Au}} < v_{2}^{\textrm{U}}$. Its magnitude increases with increase in the system size.   

\subsection{Model comparison}
\label{ssec:v2ampt}
The AMPT model is a hybrid Monte Carlo event generator extensively used to study relativistic heavy-ion collisions~\cite{ampt1}. The colliding nuclei in AMPT are modeled according to a deformed Wood-Saxon distribution with nuclear radius given by,
\begin{equation}
R(\theta,\phi) = R_{0}\left[ 1+\beta_{2}Y_{2,0}(\theta,\phi)+\beta_{3}Y_{3,0}(\theta,\phi)\right].
\end{equation} 
$R_{0}$ represents the radius parameter, $\beta_{2}$ and $\beta_{3}$ are the quadrupole and octupole deformities, and $Y_{l,m}(\theta,\phi)$ are the spherical harmonics. We studied two different cases of Wood-Saxon parameters for Ru+Ru and Zr+Zr collisions at $\sqrt{s_{\mathrm {NN}}}$ = 200 GeV, as shown in Table~\ref{tab:ampt}~\cite{ampt2}. About 9 million minimum bias events for Ru+Ru and Zr+Zr collisions at $\sqrt{s_{\mathrm {NN}}}$ = 200 GeV with parton-parton cross-section three mb have been analyzed for each case.
\begin{table}[!htbp]
\centering
\caption{Parameter set for various deformation configurations of the Ru and Zr nuclei in the AMPT model}
\vspace{0.075in}
\label{tab:ampt}
\begin{tabular}{c|cc|cc}
\hline \rule{0pt}{14pt}
	\textbf{Parameter}	&\multicolumn{2}{c|}{\textbf{Default}}	&\multicolumn{2}{c}{\textbf{Deformed}} \\[0.5ex]
\hline \rule{0pt}{14pt}
	{System}			&{Ru} &{Zr}				&{Ru} &{Zr} \\[0.5ex]
\hline \rule{0pt}{14pt}
	{$R_{0}$}			& 5.096 & 5.096 		& 5.090 & 5.090 \\[0.5ex]
	{$a$}				& 0.540	& 0.540			& 0.460 & 0.520 \\[0.5ex]
	{$\beta_{2}$}		& 0.000	& 0.000			& 0.162 & 0.060 \\[0.5ex]
	{$\beta_{3}$}		& 0.000	& 0.000			& 0.000 & 0.200 \\[0.5ex]
\hline 
\end{tabular}
\end{table}

\begin{figure}[!htbp]
\centering
\includegraphics[width=5cm]{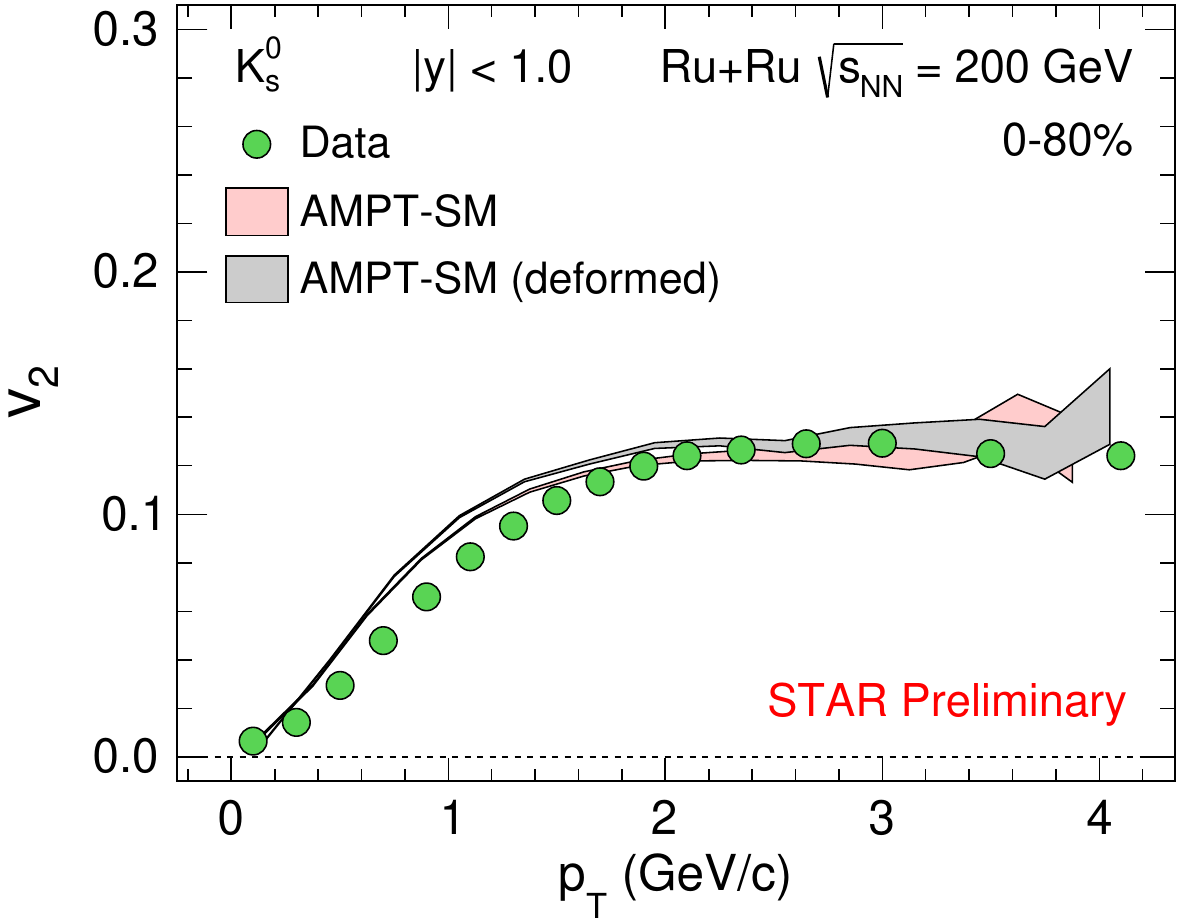}
\includegraphics[width=5cm]{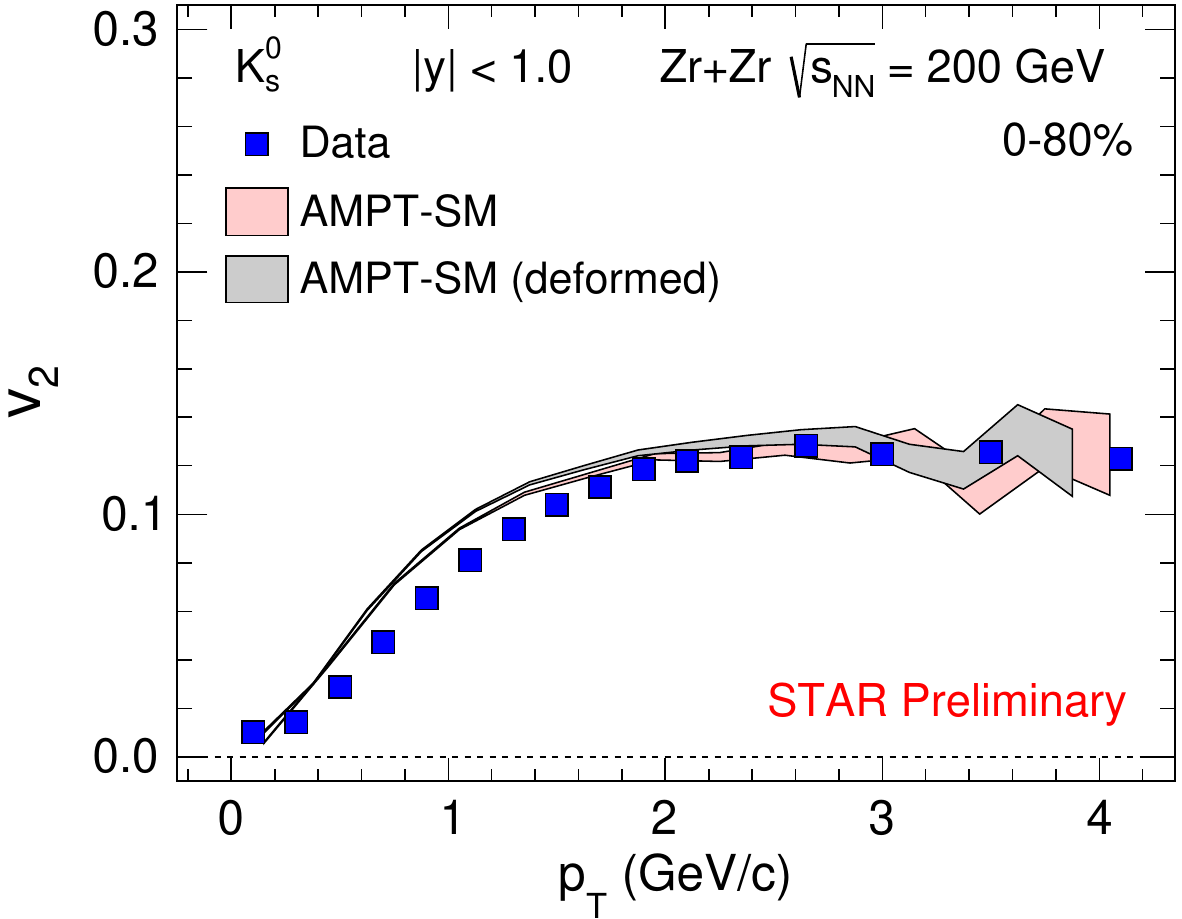}

\includegraphics[width=5cm]{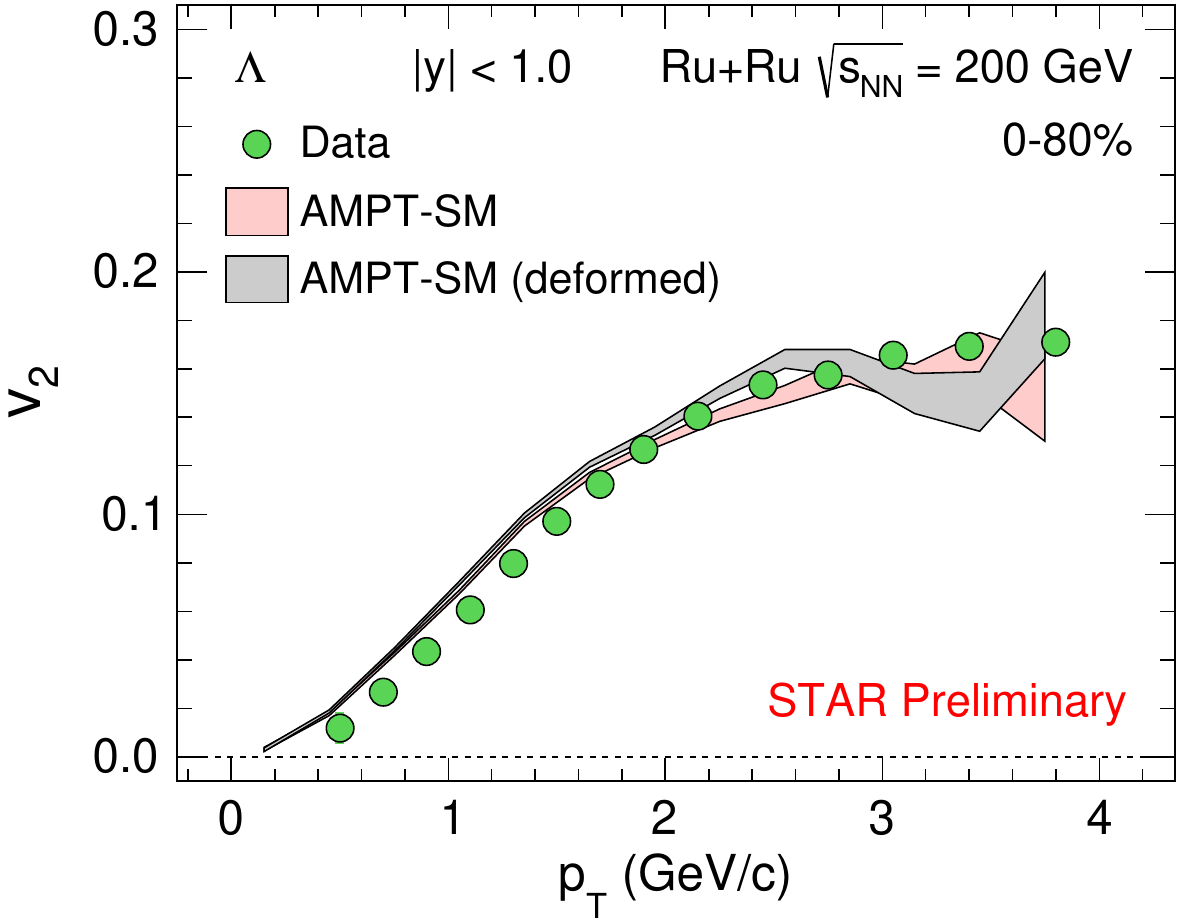}
\includegraphics[width=5cm]{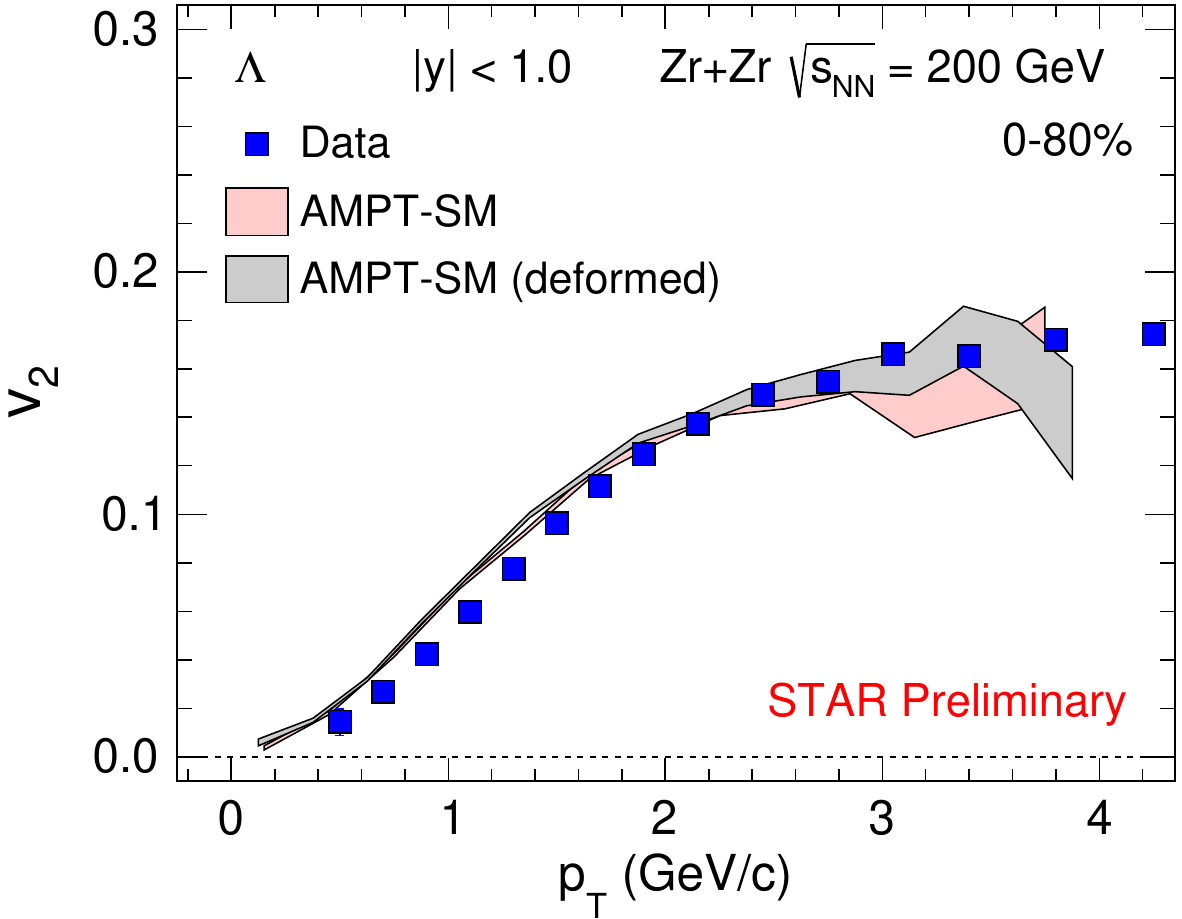}

\includegraphics[width=5cm]{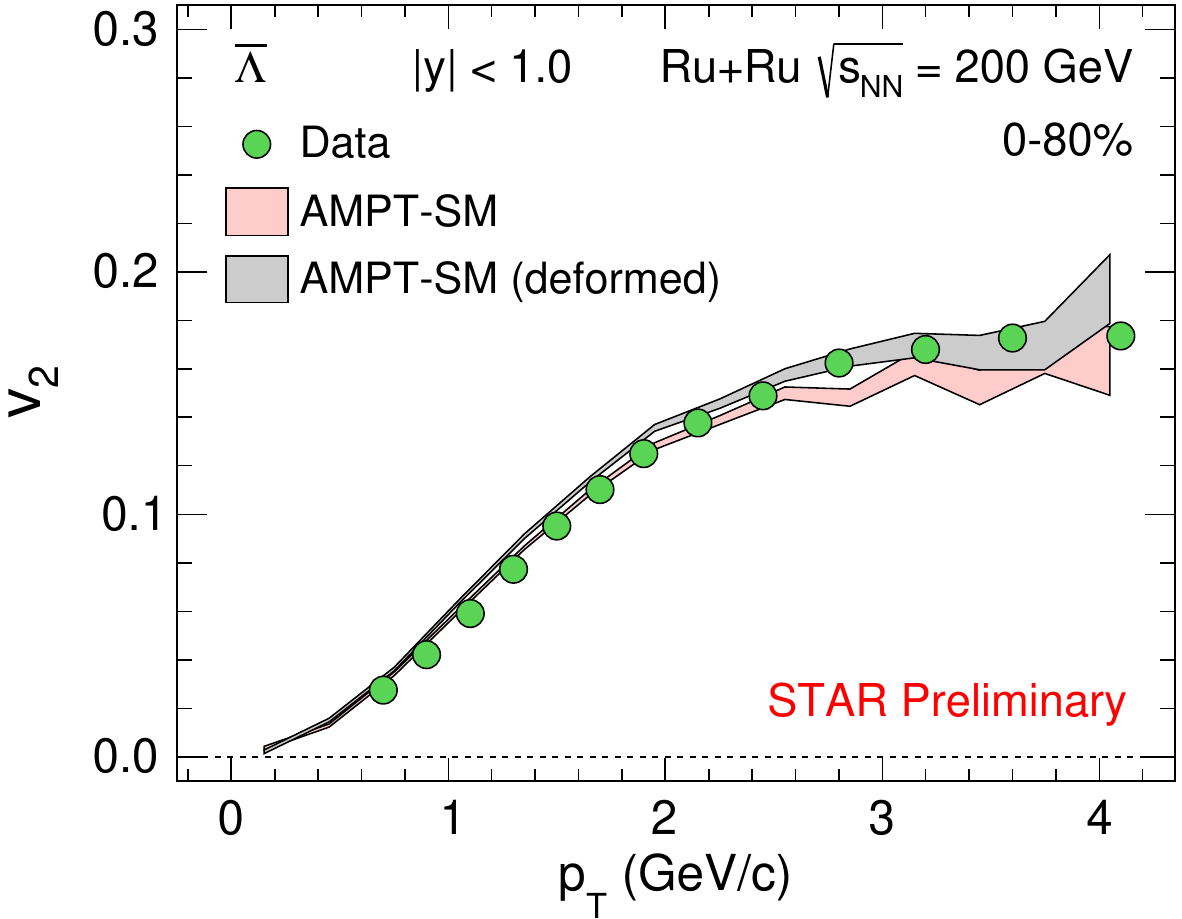}
\includegraphics[width=5cm]{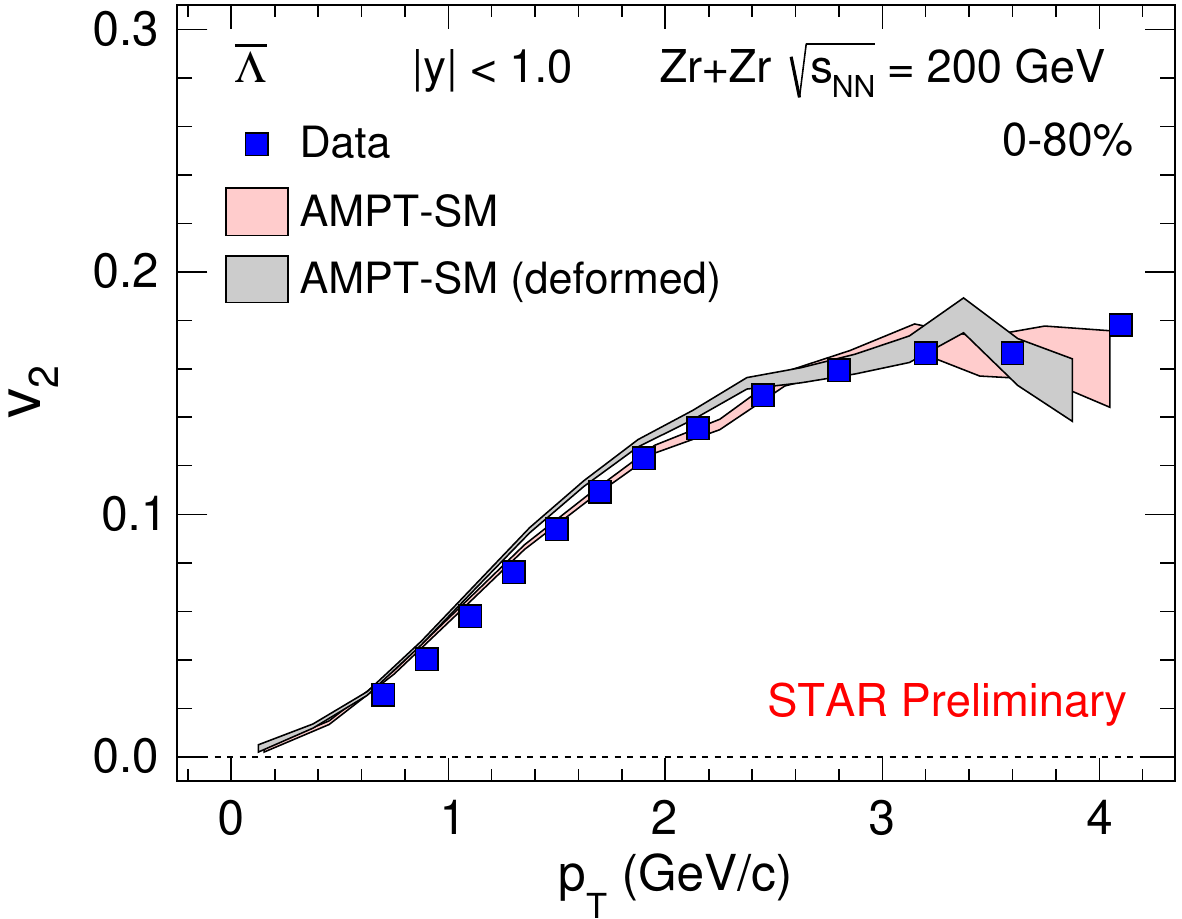}
\caption{$v_{2}$ as a function of $p_{T}$ for strange hadrons at mid-rapidity in minimum bias Ru+Ru and Zr+Zr collisions at $\sqrt{s_{\mathrm {NN}}}$ = 200 GeV compared to the AMPT model calculations~\cite{ampt1,ampt2}. The error bars represent statistical and systematic uncertainties added in quadrature.}
\label{fig:v2amptru}
\end{figure}

\begin{figure}[!htbp]
\centering
\includegraphics[width=5cm]{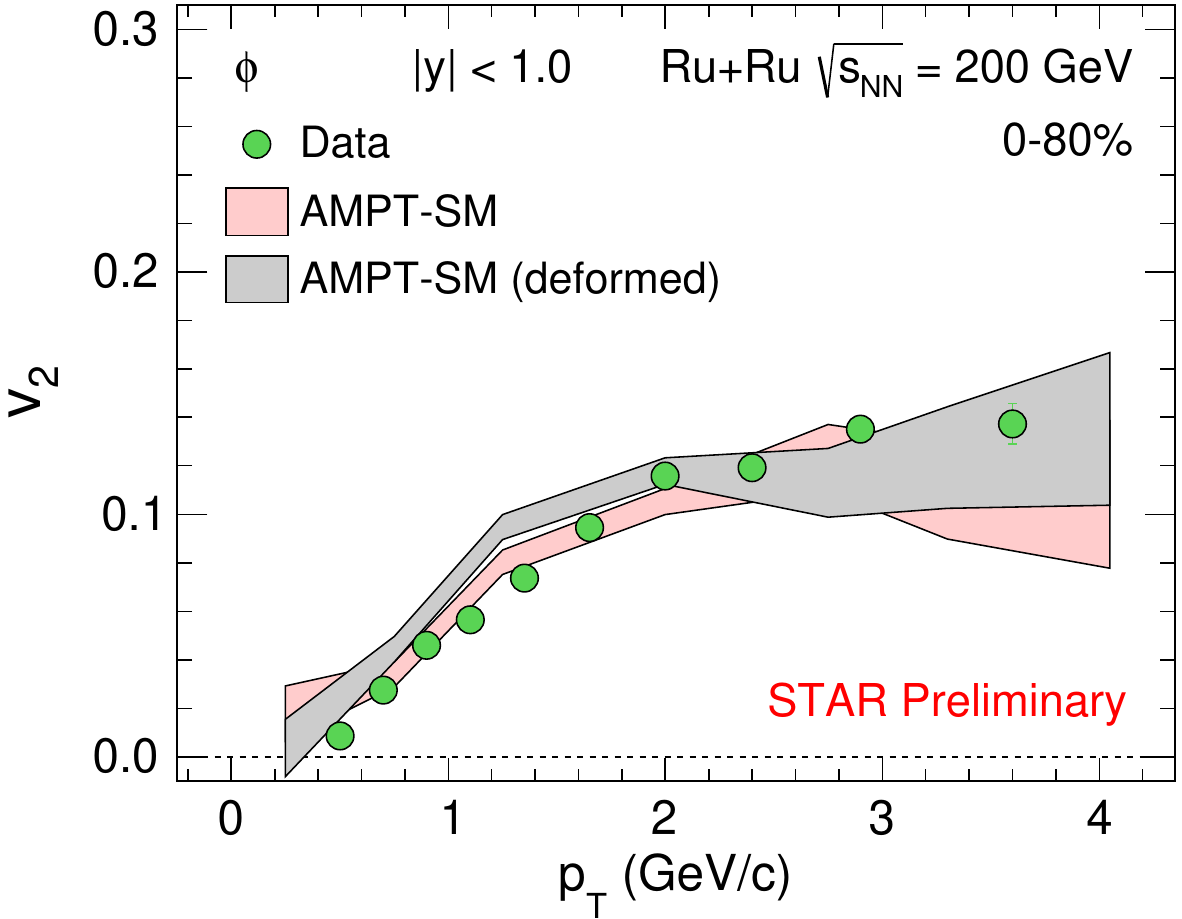} 
\includegraphics[width=5cm]{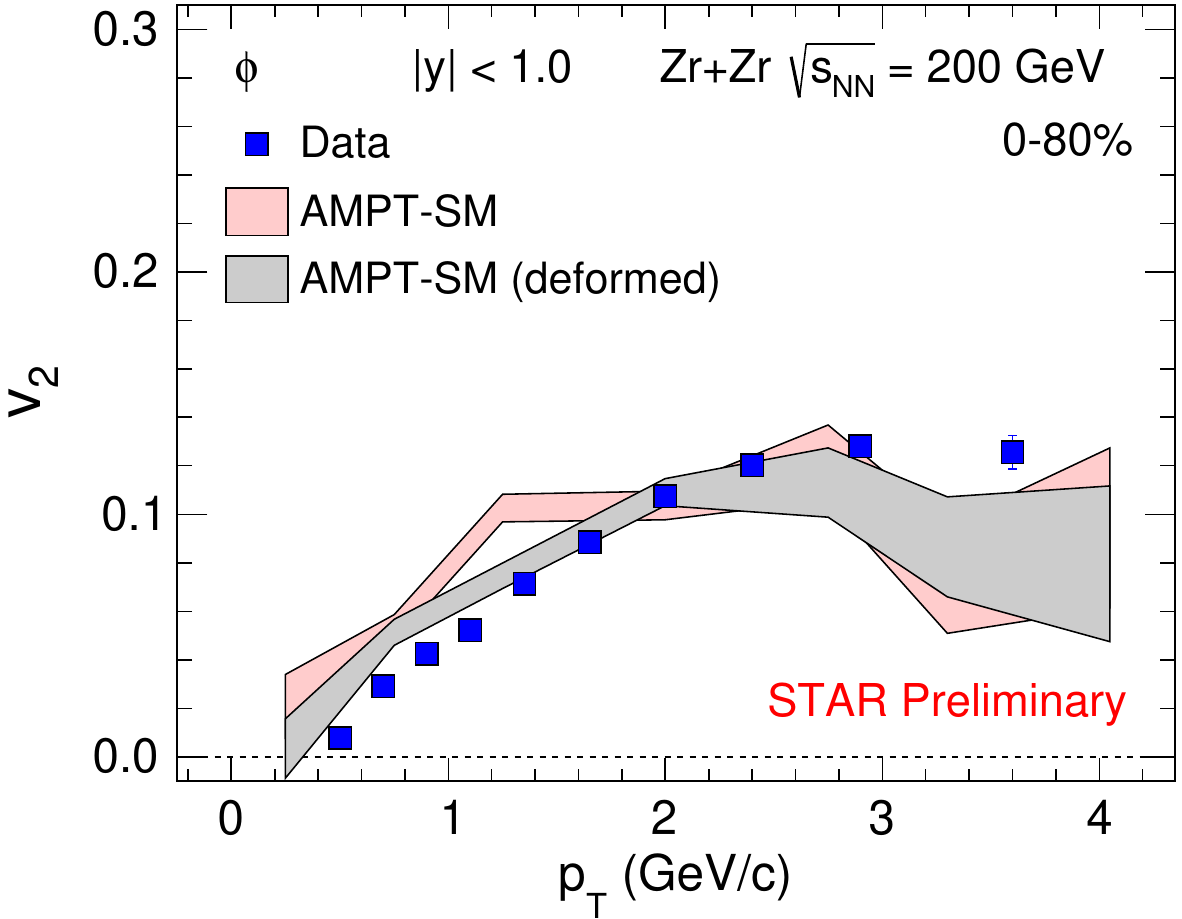} 

\includegraphics[width=5cm]{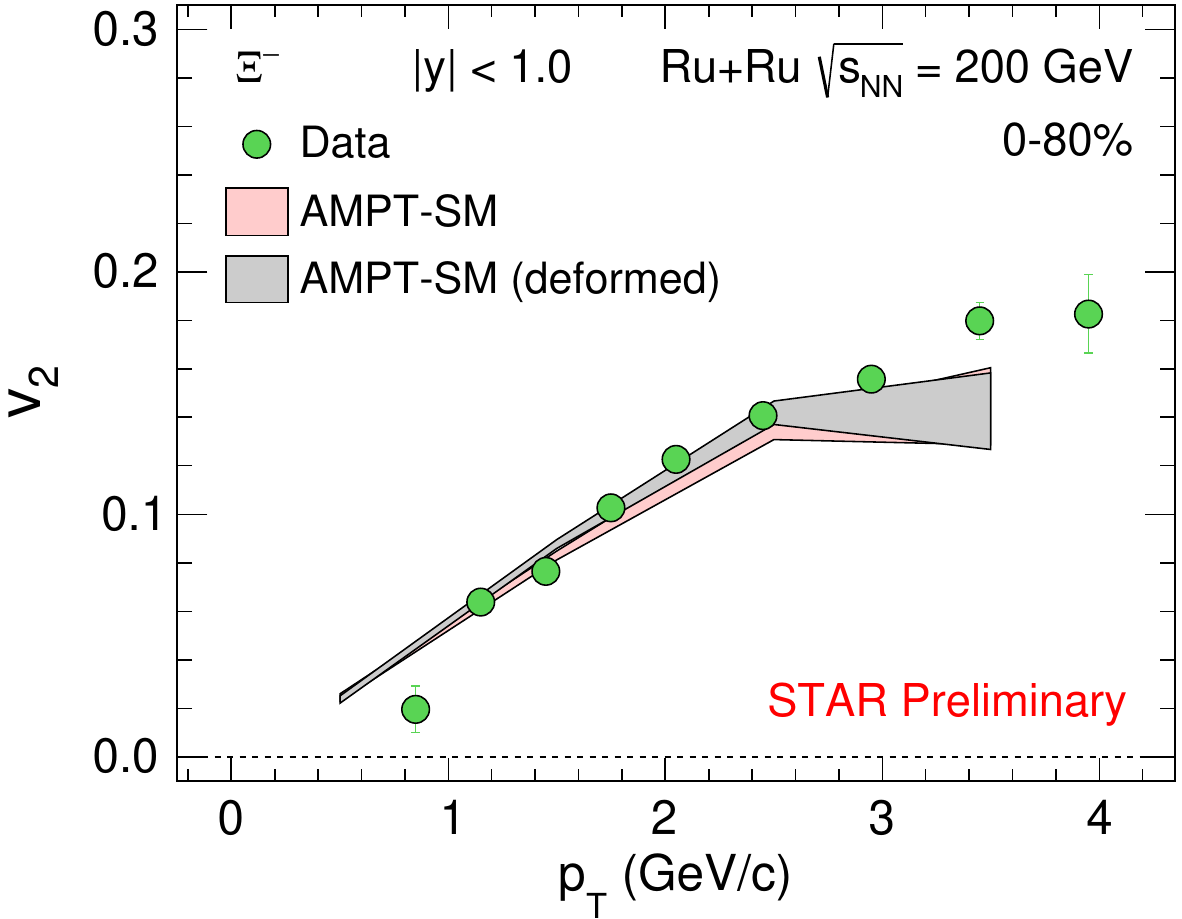} 
\includegraphics[width=5cm]{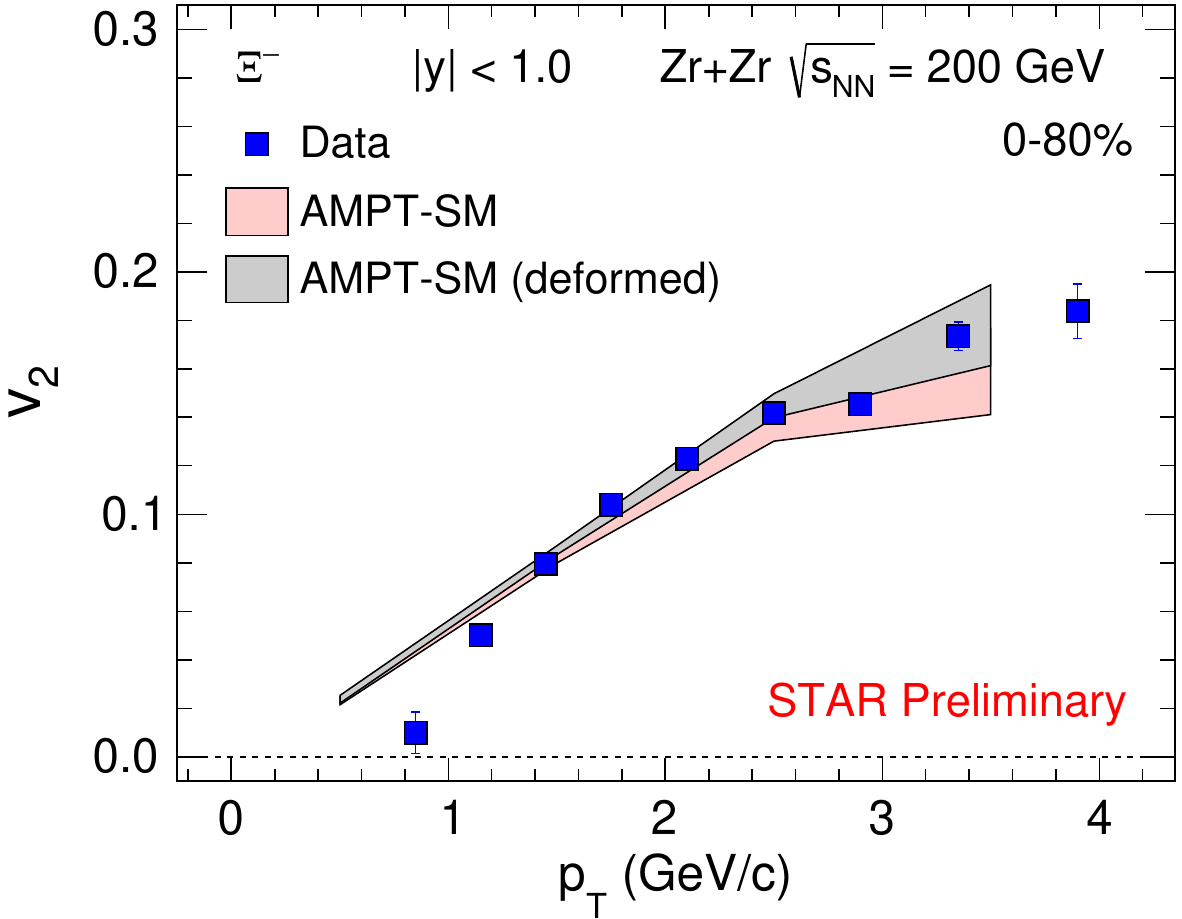} 

\includegraphics[width=5cm]{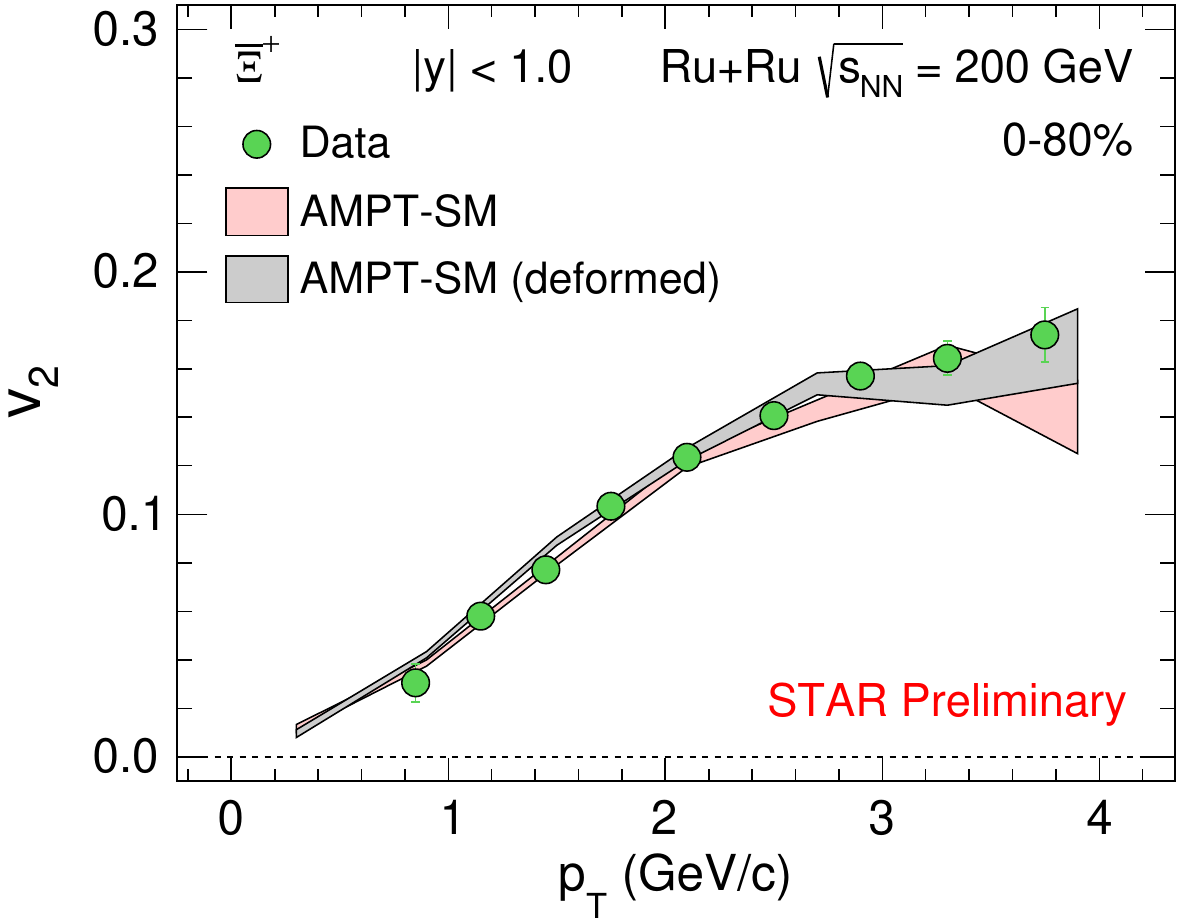} 
\includegraphics[width=5cm]{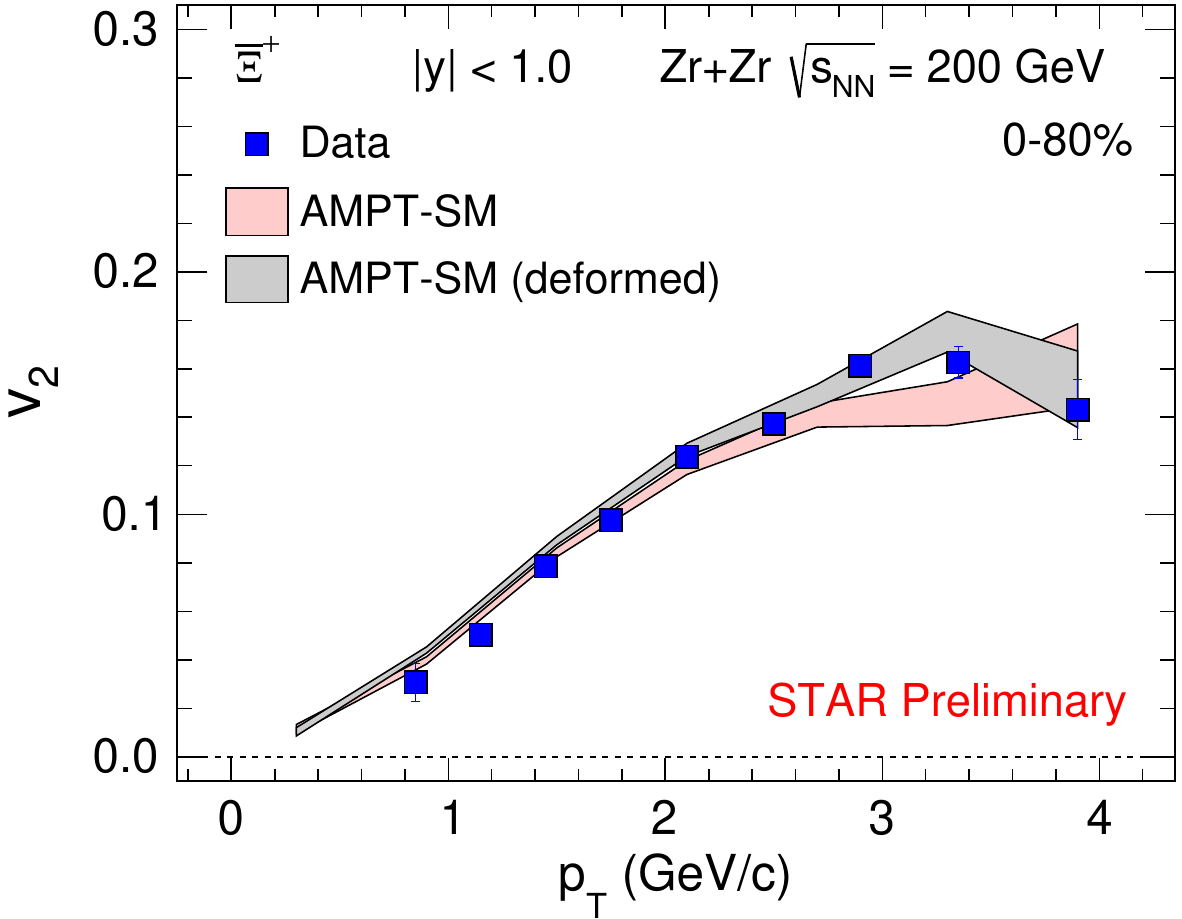} 

\includegraphics[width=5cm]{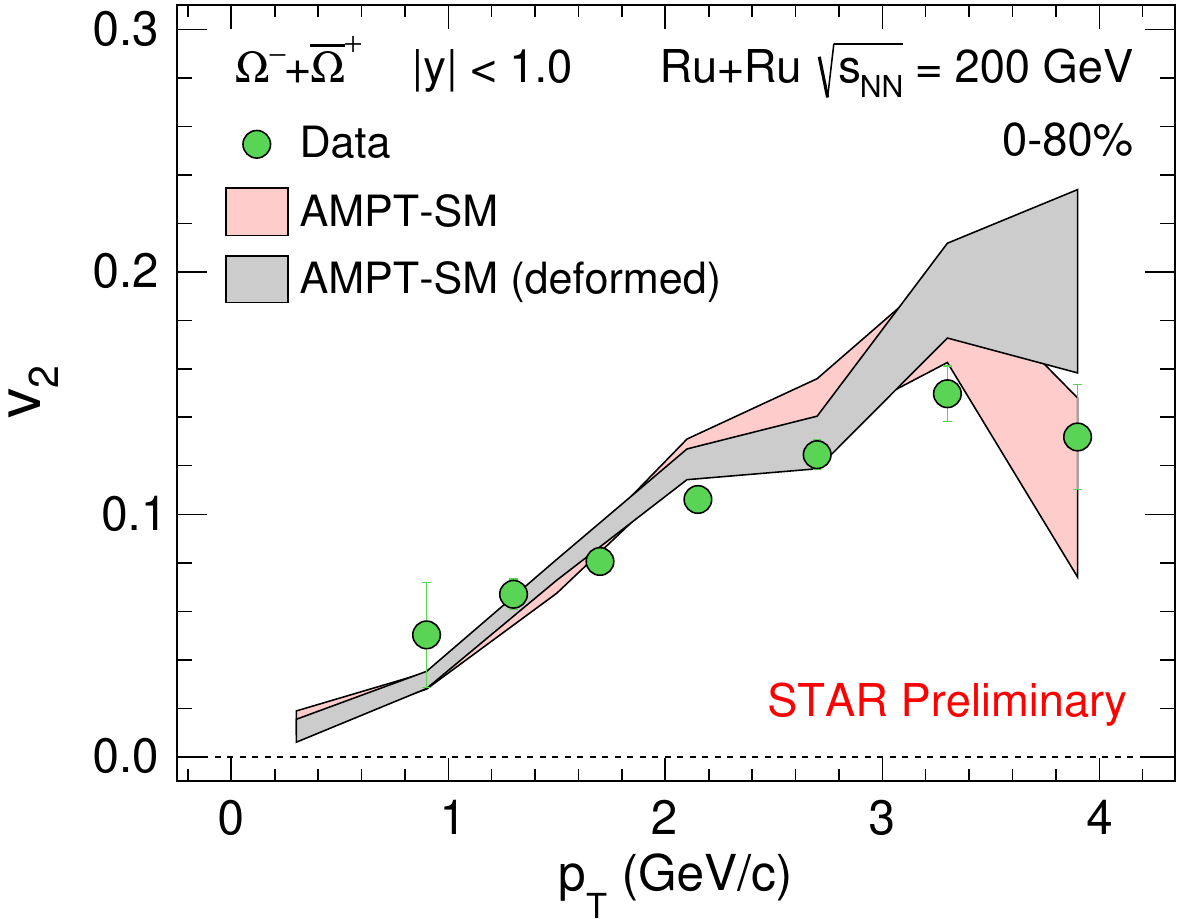}
\includegraphics[width=5cm]{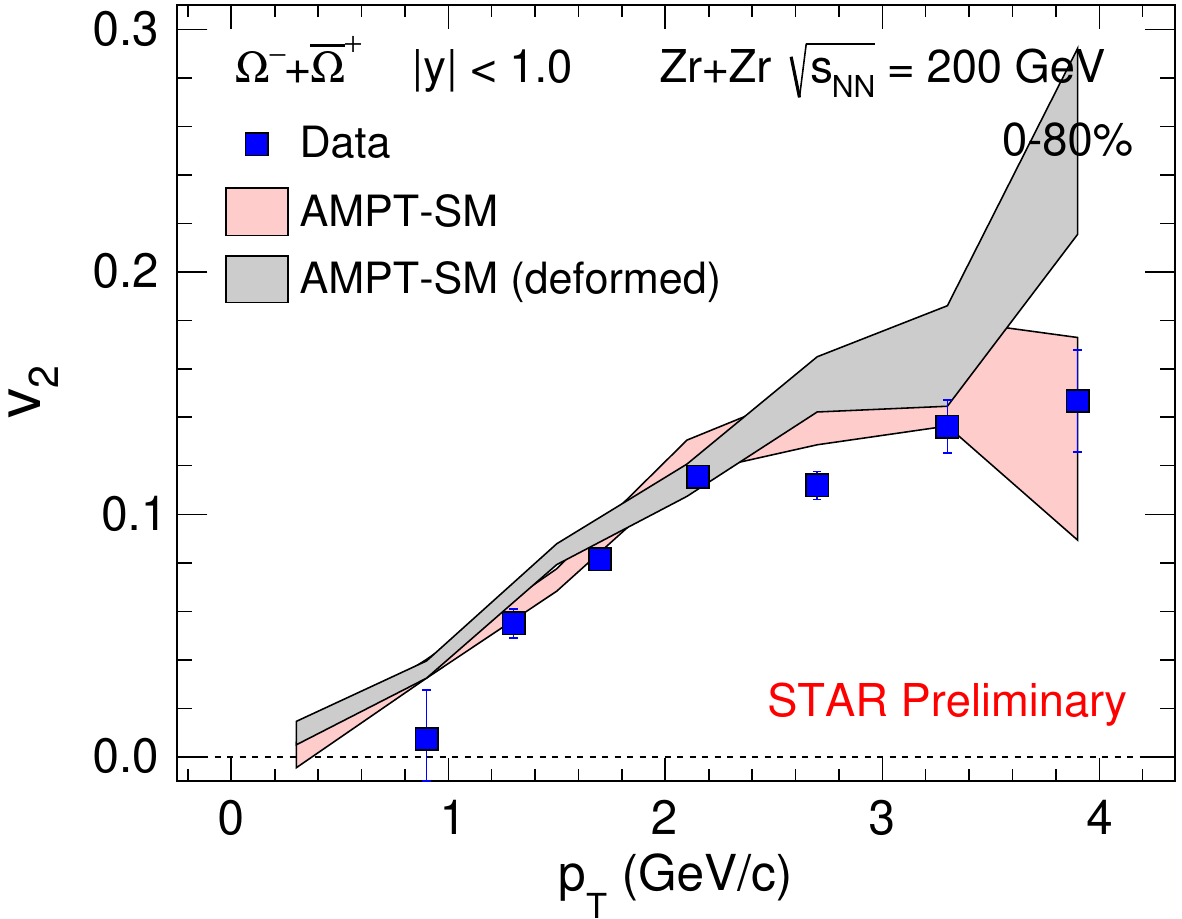}
\caption{$v_{2}$ as a function of $p_{T}$ for multi-strange hadrons at mid-rapidity in minimum bias Ru+Ru and Zr+Zr collisions at $\sqrt{s_{\mathrm {NN}}}$ = 200 GeV compared to the AMPT model calculations~\cite{ampt1,ampt2}. The error bars represent statistical and systematic uncertainties added in quadrature.}
\label{fig:v2amptzr}
\end{figure}

Figure~\ref{fig:v2amptru} and~\ref{fig:v2amptzr} show $v_{2}$ of strange and multi-strange hadrons in minimum bias Ru+Ru and Zr+Zr collisions at $\sqrt{s_{\mathrm {NN}}}$ = 200 GeV compared to the AMPT-SM model calculations. AMPT-SM model with and without nuclear deformation are close to each other, and the data in the measured $p_{T}$ range for minimum-bias isobar collisions at $\sqrt{s_{\mathrm {NN}}}$ = 200 GeV.

\section{Summary}
\label{summary}
We reported transverse momentum dependence of elliptic flow of $K_{s}^{0}$, $\Lambda$, $\bar{\Lambda}$, $\phi$, $\Xi^{-}$, $\overline{\Xi}^{+}$, and $\Omega^{-}$+$\overline{\Omega}^{+}$ at mid-rapidity in Ru+Ru and Zr+Zr collisions at $\sqrt{s_{\mathrm {NN}}}$ = 200 GeV for minimum bias (0-80\%) and in three centrality intervals (0-10\%, 10-40\%, and 40-80\%). A clear centrality dependence of $v_{2}$ is observed in the isobar collisions. We observed a particle mass hierarchy of $v_{2}$, which suggests hydrodynamic behavior at low $p_{T}$. A baryon-meson splitting of $v_{2}$ at intermediate $p_{T}$ is also observed. The elliptic flow of strange and multi-strange hadrons follows the number of constituent quark scaling, further indicating quark coalescence as the dominant particle production mechanism and the collectivity of the medium. The ratio of $p_{T}$-integrated $v_{2}$ for strange hadrons between the two isobar collisions shows a deviation from unity, which indicates different intrinsic nuclear structures of the two isobars. We observed a system size dependence of the $v_{2}$. The AMPT-SM model, with and without nuclear deformation, provides a good description of the data within the measured $p_{T}$ range for minimum-bias isobar collisions at $\sqrt{s_{\mathrm {NN}}}$ = 200 GeV. These measurements are helpful to shed light on the effect of deformation and collision geometry on anisotropic flow of particle production in relativistic heavy-ion collisions.

\end{document}